\documentclass[aps,pre,twocolumn,reprint,preprintnumbers,floatfix,nofootinbib]{revtex4}

\usepackage{amsmath}
\usepackage{amssymb}

\usepackage{graphicx}
\usepackage{dcolumn}
\usepackage{bm}
\usepackage{marvosym}

\usepackage{color}
\definecolor{light-gray}{gray}{0.5}
\definecolor{blue}{rgb}{0.0,0.0,1.0}
\definecolor{green}{rgb}{0.0,0.5,0.0}
\definecolor{red}{rgb}{1.0,0.0,0.0}
\definecolor{cyan}{rgb}{0.0,0.75,0.75}
\definecolor{magenta}{rgb}{0.75,0.0,0.75}
\definecolor{yellow}{rgb}{0.75,0.75,0.0}

\newcommand{\avg}[1]{\langle{#1}\rangle}
\newcommand{\sdot}{\cdot}
\newcommand{\ox}{\otimes}
\newcommand{\grad}{\bm \nabla}
\newcommand{\pd}{\partial}
\newcommand{\lrbig}[1]{\left( {#1} \right)}

\newcommand{\etal}{\emph{et al. }}

\newcommand{\lt}{\left}
\newcommand{\rt}{\right}

\newcommand{\D}{\Delta}
\newcommand{\dd}{\mathrm{d}}

\begin{document}
\title{Strong polymer-turbulence interactions in viscoelastic turbulent channel flow}
\author{V. Dallas}
\email{vassilios.dallas04@imperial.ac.uk}
\author{J. C. Vassilicos}
\email{j.c.vassilicos@imperial.ac.uk}
\affiliation{Institute for Mathematical Sciences, Imperial College, London, SW7 2PG, UK}
\affiliation{Department of Aeronautics, Imperial College, London, SW7 2AZ, UK}
\author{G. F. Hewitt}
\affiliation{Department of Chemical Engineering and Chemical Technology, Imperial College, London, SW7 2AZ, UK}
\begin{abstract}
This paper is focused on the fundamental mechanism(s) of viscoelastic turbulence that lead to polymer induced turbulent drag reduction phenomenon. A great challenge in this problem is the computation of viscoelastic turbulent flows, since the understanding of polymer physics is restricted to mechanical models. An effective state-of-the-art numerical method 
to solve the governing equation for polymers modelled as non-linear springs, without using any artificial assumptions as usual, was implemented here for the first time on a three-dimensional channel flow geometry. The capability of this algorithm to capture the strong polymer-turbulence dynamical interactions is depicted on the results, which are much closer qualitatively to experimental observations. This allowed a more detailed study of the polymer-turbulence interactions, which yields an enhanced picture on a mechanism resulting from the polymer-turbulence energy transfers.

\end{abstract}
\maketitle

\section{\label{sec:intro}Introduction}
A few parts per million by weight polymer in a wall-bounded turbulent flow are enough to reduce the force necessary to drive the flow through a channel by a factor of up to 70\%, as was discovered by Toms \cite{toms48} while performing experiments on the degradation of polymers. Turbulence is a multiscale phenomenon with a vast spectrum of spatial scales and therefore a very large number of degrees of freedom. Due to the fact that even the maximum polymer molecule end-to-end distance $L_p$ is much less than the Kolmogorov viscous scale $\eta$, one might anticipate that the small size polymers can only affect sub-Kolmogorov scale processes and that scales of length $\ell > \eta$ would remain unaffected. Surprisingly, the dynamics of the small polymer chains are able to fundamentally modify the large scale structures and statistics, as observed in the drag reduction (DR) phenomenon \cite{whitemunghal08}.

Polymer drag reduction in wall-bounded turbulent flows induces higher mean
velocities, implying deviations from the classical phenomenology of hydrodynamic wall-bounded turbulence and hence from the von K\'arm\'an law
\begin{equation}
 \label{eq:loglaw}
 U_+ = \frac{1}{\kappa}\ln y_+ + B
\end{equation}
where `$+$' denotes normalisation with the friction velocity $u_\tau$ and the viscous length scale $\delta_\nu \equiv \nu/u_\tau$ with $\nu$ being the fluid's kinematic viscosity. Moreover, $\kappa$ in Eq. \eqref{eq:loglaw} is the von K\'arm\'an coefficient \cite{nagibchauhan08,dvh09}, which is usually considered to be a constant taking the value $0.41$ and $B \simeq 5.2$ is the intersept constant. The detailed experimental work by Warholic \etal \cite{warholicetal99} distinguished between polymer induced drag reduced flows at low drag reduction (LDR) and high drag reduction (HDR) regimes, based on the statistical trends of the turbulent velocity field. When $|\text{DR}| \lesssim 40\%$ (LDR), the mean velocity profile is a log-law parallel to the von K\'arm\'an law Eq. \eqref{eq:loglaw} with a higher value of $B$, i.e. larger mean velocity. However, for $40\% < |\text{DR}| \lesssim 60\%$ (HDR), the slope of the log-region increases until it reaches the empirical maximum drag reduction (MDR) asymptote
\begin{equation}
 \label{eq:MDRlaw} 
 U_+ = \frac{1}{\kappa_\mathrm{v}}\ln y_+ + B_\mathrm{v}
\end{equation}
where $\kappa_\mathrm{v}^{-1} \simeq 11.7$ and the intercept constant $B_\mathrm{v} \simeq 17$. This universal asymptotic profile was discovered experimentally in pipe flow by Virk \etal \cite{virketal67,virk75} and confirmed experimentally in channel flow by Warholic \etal \cite{warholicetal99}. Overall, the mean velocity profile is bounded between the von K\'arm\'an law Eq. \eqref{eq:loglaw} and the MDR law Eq. \eqref{eq:MDRlaw}, the latter being independent of the Newtonian solvent, the polymer characteristics and the flow geometry.

Polymer induced drag reduction has been known for more than sixty years and has attracted attention both from the fundamental and applied perspective. However, no generally accepted theory has been provided to explain adequately the phenomenon. Such a theory should provide an explanation of the drag reduction onset, as well as the MDR law and its universality, which plays a significant fundamental role in understanding the phenomenon. Several theoretical concepts have been proposed but all have been subjected to criticism. The proposed theories fall mainly into two categories, that of viscous \citep{lumley69,procacciaetal08} and that of elastic effects \citep{tabordegennes86,joseph90,sreeniwhite00}.

Recent progress in DNS of viscoelastic turbulence has begun to elucidate some of the dynamical interactions between polymers and turbulence, which are responsible for drag reduction. The aim of this study is to investigate the polymer dynamics, their influence on flow quantities and the phenomenology of drag reduction in the various regimes
through DNS of viscoelastic turbulent channel flow using the finite extensible non-linear elastic  model with the Peterlin linearisation (FENE-P) \citep{birdetal87}, the most widely used coarse-grained model in such studies.

The paper is organised as follows. The necessary details on the DNS of viscoelastic turbulent channel flow are provided in section \ref{sec:dnsp}. We analyse various viscoelastic turbulent statistics in section \ref{sec:viscoelasticstats} for all the drag reduction regimes achieved in this study with a state-of-the-art numerical approach, which we have adapted to wall-bounded flows \cite{myphdthesis}, aimed at capturing discontinuities in the polymer field. This approach is described in some detail in appendix \ref{sec:fenepsolver}. Specifically, the effects of polymer extensibility and Reynolds number are briefly considered, whereas the statistics of mean velocity, fluctuating velocities and vorticities are examined in depth demonstrating that our computations are qualitatively closer to experimental observations than previous numerical studies. Section \ref{sec:conftensor} presents the conformation tensor statistics and the scaling of polymer stress tensor components at the high Weissenberg number limit, which assists in a new asymptotic result for the shear stress balance (see section \ref{sec:stressbalance}). Finally, the polymer-turbulence interactions are studied in section \ref{sec:energybalance} through the energy balance. A refined and extended picture of a conceptual model for drag reduction based on viscoelastic dissipation is proposed in section \ref{sec:DRmechanism} before summing up our most important results (see section \ref{sec:conclusion}).

\section{DNS of viscoelastic turbulent channel flow}
\label{sec:dnsp}

The enormous number of degrees of freedom of each coil means that polymers are an extraordinarily complex system, whose dynamics depend on the conformations of the polymer molecules, i.e. orientation and degree of stretching of a coil. The study of detailed motions of this complex system and their relations to the non-equilibrium properties would be prohibitive. Only after elimination of the fast relaxation processes of local motions in favour of stochastic noise, is it possible to study the dynamics of longer relaxation time scales \citep{ottinger96}, such as the end-to-end conformation, that are responsible for many physical properties of polymers in fluids, such as viscoelastic turbulence and polymer drag reduction. Thus, coarse-grained mechanical models, such as bead-rod-spring models, are very crucial in DNS of viscoelastic turbulence.

The computationally demanding Navier-Stokes equations in three-dimensions makes a Lagrangian approach for the polymer equally prohibitive and also limits polymer models to simple representations. A successful model for DNS studies of turbulent drag reduction is the FENE-P model in the Eulerian frame of reference, representing a conformation field of polymer macromolecules that have been modelled as non-linear bead-spring dumbbells \cite{birdetal87}. The standard approach to numerically solve the FENE-P model and its slight variations \cite{sureshkumarberis95,minetal01} add an artificial diffusion term in the conformation field equation to avoid the loss of strict positive definiteness (SPD) of the conformation tensor and subsequently numerical breakdown caused by the hyperbolic nature of the FENE-P model (see Eq. \eqref{eq:nondimfenep}).

Jin and Collins \cite{jincollins07} stress the fact that much finer grid resolutions are required to fully resolve the polymer field than the velocity and pressure fields. Indeed, the hyperbolicity of the FENE-P model admits near discontinuities in the conformation and polymer stress fields \cite{josephsaut86}. Qualitatively similar problems occur with shock waves and their full resolution in gas dynamic compressible flows, which is not practical using finer grids. In this case, high resolution numerical schemes such as slope-limiter and Godunov-type methods \cite{leveque02} have proved successful at capturing the shock waves by accurately reproducing the Rankine-Hugoniot conditions across the discontinuity to ensure the correct propagation speed. 

Motivated by these schemes, Vaithianathan \etal \cite{vaithietal06} adapted the second-order hyperbolic solver by Kurganov and Tadmor \cite{kt00}, which guarantees that a positive scalar remains positive over all space, to satisfy the SPD property of the conformation tensor and therefore avoid loss of evolution. Vaithianathan \etal \cite{vaithietal06} further demonstrated that this scheme dissipates less elastic energy than methods based on artificial diffusion, resulting in stronger polymer-turbulence interactions. Moreover, according to the most recent review on the subject \cite{whitemunghal08}, there are a lot of divergent and misleading results because of this artificial term introduced in the governing equations. For these reasons a modification of this shock-capturing scheme was developed in this present study to comply with non-periodic boundary conditions (see appendix \ref{sec:fenepsolver}). 

\subsection{Governing equations}
The dimensionless incompressible Navier-Stokes equations for a viscoelastic fluid take the form
\begin{equation}
 \begin{gathered}
  \label{eq:nondimNS2}
  \grad \sdot \bm u = 0 \\
  \pd_t \bm u
  + (\bm u \sdot \grad) \bm u
  = - \grad p 
  + \frac{\beta}{\text{Re}_c} \bm{\D u}
  + \grad \sdot \bm \sigma
 \end{gathered}
\end{equation}
where 
$\beta \equiv \mu_s / \mu_0$ is the ratio of the solvent viscosity $\mu_s$ to the total zero-shear-rate viscosity of the solution $\mu_0$, $\text{Re}_c \equiv U_c \delta / \nu$ is the Reynolds number based on $U_c \equiv \frac{3}{2}U_b$ with $U_b$ the bulk velocity of the flow kept constant in time and the channel's half-width $\delta$. The extra force in Eq. \eqref{eq:nondimNS2} arises due to polymers and the polymer stress tensor for the FENE-P dumbbells is defined by the Kramers expression
\begin{equation}
 \label{eq:nondimpolystress}
 \bm \sigma = \frac{1-\beta}{\text{Re}_c\text{We}_c} \lrbig{f(tr\bm C)\bm C - \bm I}
\end{equation}
where $\text{We}_c \equiv \tau_p U_c / \delta$ with $\tau_p$ the polymer relaxation time scale, $f(tr\bm C) \equiv \frac{L^2_p - 3}{L^2_p - tr\bm C}$ is the Peterlin function \cite{peterlin61} and $\bm C \equiv \avg{\bm{QQ}}$ is the conformation tensor, which is defined as the dyadic product of the end-to-end vector $\bm Q$ of a dumbbell that specifies its configuration. The Peterlin function prevents the dumbbell to reach its maximum extensibility, i.e. $tr\bm C \leq L_p^2$, since as $tr\bm C \to L_p^2$ the force required for further extension approached infinity. Note that $\bm C$ and $L_p^2$ are made dimensionless by the equilibrium length scale $\sqrt{k_B T / H}$, where $k_B$ is the Boltzmann constant, $T$ is the solution temperature and $H$ is the Hookean spring constant and they have been normalised such that the equibrium condition is $\bm C_{eq} = \bm I$. Then, the conformation tensor is governed by the FENE-P model
\begin{equation}
 \label{eq:nondimfenep}
 \pd_t \bm C + (\bm u \sdot \grad)\bm C
 - \bm C \sdot \grad\bm u - \grad\bm u^\top \sdot \bm C =
 - \frac{1}{\text{We}_c}(f(tr\bm C)\bm C - \bm I)
\end{equation}
where the left hand side is the material derivative for a tensor field preserving its Galilean invariance and the right hand side represents deviation from the isotropic equilibrium due to Warner's finite extensible non-linear elastic spring-force law \cite{warner72}.

The elastic potential energy per unit volume $E_p$ stored by FENE-P dumbbells can now be specified by taking the integral of Warner's spring-force law over the end-to-end vector and after some algebra we obtain
\begin{align}
 \label{eq:elasticenergy}
 E_p = \frac{1}{2}\frac{(1-\beta)}{\text{Re}_c\text{We}_c}(L_p^2 - 3) \ln(f(tr\bm C)) + E_{p_0}
\end{align}
where $E_{p_0}$ is a constant reference energy at equilibrium. After that, taking the time derivative of the elastic potential energy
\begin{equation}
 \label{eq:dtEp}
 \pd_t E_p = \frac{1}{2}\frac{(1-\beta)}{\text{Re}_c\text{We}_c} (L_p^2 - 3) \frac{1}{f} \frac{\pd f}{\pd C_{ii}} \frac{\pd C_{ii}}{\pd t} = \frac{1}{2}\frac{(1-\beta)}{\text{Re}_c\text{We}_c} f \frac{\pd C_{ii}}{\pd t},
\end{equation}
using the trace of Eq. \eqref{eq:nondimfenep}, viz. 
\begin{equation}
 \frac{\pd C_{ii}}{\pd t} = 2 C_{ik} \pd_k u_i - \frac{1}{\text{We}_c}( f(C_{kk}) C_{ii} - \delta_{ii} )
\end{equation}
and similarly for the $\grad E_p$, we can derive the following balance equation for the elastic potential energy of FENE-P dumbbells
\begin{equation}
 \label{eq:Epeqn}
 \pd_t E_p + \bm u \sdot \grad E_p = \bm \sigma \sdot \grad \bm u - \frac{1}{2 \text{We}_c}f(tr\bm C) tr\bm \sigma
\end{equation}
where $E_p$ is produced by $\bm \sigma \sdot \grad \bm u$, dissipated by $\frac{1}{2 \text{We}_c}f(tr\bm C) tr\bm \sigma$ and transported by $\bm u \sdot \grad E_p$.

\subsection{Numerical parameters and procedures}
\label{sec:parameters}
Incompressible viscoelastic turbulence in a channel was simulated in a rectangular geometry by numerically solving the non-dimensional Eqs. \eqref{eq:nondimNS2}-\eqref{eq:nondimfenep} in Cartesian co-ordinates. After obtaining the new update of the conformation tensor from the FENE-P model using the method described in appendix \ref{sec:fenepsolver}, Eqs. \eqref{eq:nondimNS2}-\eqref{eq:nondimfenep} are numerically integrated with a fractional step method using a second-order Adams-Bashworth/Trapezoidal scheme (see appendix \ref{sec:timeadvance}). The fractional step method projects the velocity field to a divergence free velocity field and the Poisson pressure equation is solved in Fourier space with a staggered grid for the pressure field \cite{laizetlamballais09}. The staggered grid for the pressure was used for numerical stability purposes as well as the skew-symmetric implementation of the non-linear term in Eqs. \eqref{eq:nondimNS2}. Spatial derivatives are estimated using sixth-order compact finite-difference schemes \cite{lele92}. The grid stretching technique used in the inhomogeneous wall-normal direction maps an equally spaced co-ordinate in the computational space to a non-equally spaced co-ordinate in the physical space, in order to be able to use Fourier transforms \cite{laizetlamballais09, cainetal84}. Further details of our numerical method are provided in \cite{myphdthesis}. Moreover, a validation of the algorithm just for the Navier-Stokes equations for turbulent channel flow can be found in \cite{laizetlamballais09}, where this methodology was compared with spectral and second-order finite-difference schemes showing the necessity of spectral-like accuracy of the compact high-order schemes against second-order finite-differences in turbulence computations.

To simulate incompressible channel flow turbulence we applied periodic boundary conditions for $\bm u \equiv (u,v,w)$ in the $x$ and $z$ homogeneous directions and no-slip boundary conditions $\bm u = 0$ at the walls. The mean flow is in the $x$ direction, i.e. $\avg{\bm u} = (\avg{u(y)},0,0)$, where $\avg\;$ in this paper denotes averages in $x$, $z$ spatial directions and time. The bulk velocity $U_b$ in the $x$ direction was kept constant for all computations at all times by adjusting the mean pressure gradient $- \dd\avg p / \dd x$ at each time step. The choice of $U_b$ in the computations for the Newtonian fluid is made based on Dean's formula $\text{Re}_{\tau_0} \simeq 0.119 \text{Re}_{c}^{7/8}$ \cite{dean78,lesieur97} for a required $\text{Re}_{\tau_0} \equiv \frac{u_{\tau_0} \delta}{\nu}$, where $u_{\tau_0}$ is the friction velocity for Newtonian fluid flow, i.e. $\beta = 1$ (see N cases in Table \ref{tbl:dnspparameters}).

The procedure used for the computation of the viscoelastic turbulent channel flows of Table \ref{tbl:dnspparameters} is the following. First, DNS of the Newtonian fluid, i.e. $\beta = 1$, were performed for the various Reynolds numbers until they reached a steady state. Then, the initial conditions for the viscoelastic DNS were these turbulent Newtonian velocity fields as well as the stationary analytical solution of the FENE-P model, given a steady unidirectional shear flow $\bm u = (U(y),0,0)$, for the $C_{ij}$ tensor components
\begin{align}
 \label{eq:fenepsln}
  C_{11} &= \frac{1}{f(C_{kk})}\lrbig{1 + \frac{2\text{We}_c^2}{f^2(C_{kk})} \lrbig{\frac{\dd U}{\dd y}}^2} \nonumber\\
  C_{12} &= \frac{\text{We}_c}{f^2(C_{kk})} \frac{\dd U}{\dd y} \nonumber\\
  C_{13} &= C_{23} = 0 \nonumber\\
  C_{22} &= C_{33} = \frac{1}{f(C_{kk})} \nonumber\\
  f(C_{kk}) &= \frac{2}{3}\cosh\frac{\phi}{3} + \frac{1}{3}
\end{align}
with $\phi = \cosh^{-1}\lrbig{\frac{27}{2}\Omega^2 + 1}$, $\Omega = \frac{\sqrt{2}\text{We}_c}{L_p} \frac{\dd U}{\dd y}$ and $\frac{\dd}{\dd y}U = - 6(y-1)^7$ assuming that $U(y) = 0.75(1 - (y-1)^8) \; \forall \, y \in [0,2]$ is a close approximation to the averaged velocity profile of a Newtonian fully developed turbulent channel flow at moderate Reynolds numbers \cite{moinkim80}. Initially, the governing equations were integrated uncoupled, i.e. $\beta = 1$, until the conformation tensor achieved a stationary state. From then on the fully coupled system of equations, i.e. $\beta \neq 1$, was marched far in time, while $\bm u$ and $\bm C$ statistics were monitored for several successive time integrals until a fully developed steady state is reached, which satisfies the total shear stress balance across the channel, viz.
\begin{equation}
 \label{eq:mombalancep}
 \frac{\beta}{\text{Re}_c}\frac{\dd \avg{u}}{\dd y} - \avg{u'v'} + \avg{\sigma_{12}} = u_\tau^2 \lrbig{1 - \frac{y}{\delta}}
\end{equation}
where $\avg{\sigma_{12}} = \frac{1 - \beta}{\text{Re}_c \text{We}_c} \avg{\frac{L_p^2 - 3}{L_p^2 - C_{kk}}C_{12}}$ is the mean polymer shear stress. Finally, after reaching a statistically steady state, statistics were collected for several decades of through-flow time scales $L_x / U_b$. In addition, existing turbulent velocity and conformation tensor fields were restarted for computations where $\text{We}_c$ or $L_p$ was modified. In these cases, the flow undergoes a transient time, where again sufficient statistics were collected after reaching a stationary state.

According to Eqs. \eqref{eq:nondimNS2}-\eqref{eq:nondimfenep}, the four dimensionless groups that can fully characterise the velocity and the conformation tensor fields are $\text{We}_c$, $L_p$, $\beta$ and $\text{Re}_c$, and they are tabulated below. ?he reasons behind the choice of the particular parameter values is outlined below. The rationale here follows the thorough parametric study by Li \etal \cite{lietal06}.
\begin{widetext}
\begin{center}
\begin{table}[!ht]
 \caption{Parameters for the DNS of viscoelastic turbulent channel flow. The friction Weissenberg number is defined by $\text{We}_{\tau_0} \equiv \tau_p u_{\tau_0}^2 / \nu$. LDR cases: A, B, D2, I, J; HDR cases: C, D, D1, E, F, G, K; MDR case: H.}
 \label{tbl:dnspparameters}
   \begin{ruledtabular}
    \begin{tabular}{*{10}{c}}
      \textbf{Case} & $\bm{\text{We}_c}$ & $\bm{\text{We}_{\tau_0}}$ & $\bm{L_p}$ & $\bm{\beta}$ & $\bm{\text{Re}_c}$ & $\bm{\text{Re}_\tau}$ & $\bm{L_x \times L_y \times L_z}$ & $\bm{N_x \times N_y \times N_z}$ & $\bm {\text{DR}(\%)}$ \\
      \hline
      N1 &  - &   -   &   - &   1 &  2750 & 123.8 & $6.5\pi\delta \times 2\delta \times 1.5\pi\delta  $ & $200 \times  65 \times 100$  &   0   \\
      N2 &  - &   -   &   - &   1 &  4250 & 181   & $4.5\pi\delta \times 2\delta \times    \pi\delta  $ & $200 \times  97 \times 100$  &   0   \\
      N3 &  - &   -   &   - &   1 & 10400 & 392.6 & $2  \pi\delta \times 2\delta \times 0.5\pi\delta  $ & $200 \times 193 \times 100$  &   0   \\
      A  &  2 &  15.4 & 120 & 0.9 &  4250 & 167.7 & $4.5\pi\delta \times 2\delta \times    \pi\delta  $ & $200 \times  97 \times 100$  & -14.2 \\
      B  &  4 &  30.8 & 120 & 0.9 &  4250 & 147.3 & $4.5\pi\delta \times 2\delta \times    \pi\delta  $ & $200 \times  97 \times 100$  & -33.8 \\
      C  &  7 &  54   & 120 & 0.9 &  4250 & 121.8 & $4.5\pi\delta \times 2\delta \times    \pi\delta  $ & $200 \times  97 \times 100$  & -54.7 \\
      D  &  9 &  69.4 & 120 & 0.9 &  4250 & 118.3 & $4.5\pi\delta \times 2\delta \times    \pi\delta  $ & $200 \times  97 \times 100$  & -57.3 \\
      D1 &  9 &  69.4 & 60  & 0.9 &  4250 & 124.7 & $4.5\pi\delta \times 2\delta \times    \pi\delta  $ & $200 \times  97 \times 100$  & -52.5 \\
      D2 &  9 &  69.4 & 30  & 0.9 &  4250 & 150.3 & $4.5\pi\delta \times 2\delta \times    \pi\delta  $ & $200 \times  97 \times 100$  & -31   \\
      E  & 11 &  84.8 & 120 & 0.9 &  4250 & 113.3 & $4.5\pi\delta \times 2\delta \times    \pi\delta  $ & $200 \times  97 \times 100$  & -60.8 \\
      F  & 13 & 100.2 & 120 & 0.9 &  4250 & 112.4 & $4.5\pi\delta \times 2\delta \times    \pi\delta  $ & $200 \times  97 \times 100$  & -61.4 \\
      G  & 15 & 115.6 & 120 & 0.9 &  4250 & 111.4 & $4.5\pi\delta \times 2\delta \times    \pi\delta  $ & $200 \times  97 \times 100$  & -62.1 \\
      H  & 17 & 131   & 120 & 0.9 &  4250 & 107.8 & $8  \pi\delta \times 2\delta \times    \pi\delta  $ & $200 \times  97 \times 100$  & -64.5 \\
      I  &  2 &  29.6 & 120 & 0.9 & 10400 & 323.3 & $2  \pi\delta \times 2\delta \times 0.5\pi\delta  $ & $200 \times 193 \times 100$  & -32.2 \\
      J  &  4 &  22.3 & 120 & 0.9 &  2750 & 106.9 & $6.5\pi\delta \times 2\delta \times 1.5\pi\delta  $ & $200 \times  65 \times 100$  & -25.4 \\
      K  &  7 &  39   & 120 & 0.9 &  2750 &  91.1 & $6.5\pi\delta \times 2\delta \times 1.5\pi\delta  $ & $200 \times  65 \times 100$  & -45.9 \\
	 \end{tabular}
  \end{ruledtabular}
\end{table}
\end{center}
\end{widetext}

Drag reduction effects are expected to be stronger at high Weissenberg numbers. In fact, higher levels of percentage drag reduction at MDR have also been measured for higher Reynolds numbers \cite{virk75}, showing the Reynolds number dependence on drag reduction amplitude. Therefore, in this work, an extensive parametric study has been carried out by mainly varying $\text{We}_c$ for the computationally affordable $\text{Re}_c = 4250$ to determine the impact of polymer dynamics on the extent of drag reduction. The Reynolds numbers considered here, $\text{Re}_c = 2750, 4250$ and $10400$, are small in comparison to most experimental studies but fall within the range of most DNS studies of polymer induced turbulent drag reduction. Nevertheless, these Reynolds numbers are sufficiently large for the flow to be always turbulent and allow to study the dynamics of viscoelastic turbulence. Different maximum dumbbell lengths were also taken into account to check their effects for the same $\text{We}_c$ and $\text{Re}_c$. The chosen $L_p^2 = b + 3$ values are representative of real polymer molecule lengths which can be related through $b \approx N_C / (\sigma^2_{sf} N)$ where $N_C$ is the number of carbon atoms in the backbone of the polymer macromolecule, $\sigma_{sf}$ is an empirical steric factor, $N$ is the number of monomers and $b$ is the finite dumbbell extensibility and is a large number \cite{ottinger96}. Note that in the limit $b \to \infty$, the Hookean spring-force law is recovered, which governs a linear spring.

Low $\beta$ values were used in most prior DNS to achieve high levels of drag reduction, in view of the attenuation of the polymer-turbulence interactions due to the additional artificial diffusion term in the FENE-P model and their moderate Reynolds numbers, usually $\text{Re}_{\tau} \equiv \delta/\delta_\nu \leq 395$. In fact, values as low as $\beta = 0.4$ have been applied thus amplifying viscoelastic effects so as to reach the HDR regime \cite{ptasinskietal03}. However, such low $\beta$ values may lead to significant shear-thinning\footnote{the shear stress increases slower than linear $\sigma_{12} \propto S_{12}$} \cite{joseph90} unlike in experiments of polymer drag reduction. The fact that the numerical scheme applied in our study for the FENE-P model is expected to capture the strong polymer-turbulence interactions allows the value of $\beta$, which is inversely proportional to the polymer concentration, to be high, i.e. $\beta = 0.9$, representative of dilute polymer solutions used in experiments.

The box sizes $L_x \times L_y \times L_z$, where subscripts indicate the three Cartesian co-ordinates, were chosen with reference to the systematic study by Li \etal \cite{lietal06} of how the domain size influences the numerical accuracy. Specifically, they point out that long boxes are required in DNS of polymer drag reduction, particularly in the streamwise direction because of longer streamwise correlations at higher percentage DR, as opposed to the minimal flow unit \cite{jimenezmoin91} used in many earlier works. Different grid resolutions $N_x \times N_y \times N_z$ were tested for convergence. In particular, the following set of resolutions $128 \times 65 \times 64$, $200 \times 97 \times 100$ and $256 \times 129 \times 128$ were tried for $\text{Re}_{\tau_0} \simeq 180$ with the two latter giving identical mean velocity profiles and not significantly different rms velocity and vorticity profiles. Similarly, grid sensitivity tests were carried out for the other $\text{Re}_{\tau_0}$ cases. Eventually, the resolutions for each Newtonian fluid computation were validated against previously published databases for the corresponding $\text{Re}_{\tau_0}$ cases \cite{mkmdnsdata,kasagidnsdata,sotondnsdata}. Note that if the resolutions for Newtonian turbulent computations are adequately resolving the flow scales, then the same resolutions are sufficient for viscoelastic turbulent computations, since the size of vortex filaments in these flows increases while their number decreases as drag reduces \cite{whitemunghal08}.

For a given resolution, viscoelastic computations require approximately 4 times more memory and 2 times more CPU time per time step compared to the Newtonian case. The time step $\Delta t$ used in viscoelastic computations is typically a factor of $5$ smaller than that used in the Newtonian cases due to the stricter CFL condition of the present numerical method for the FENE-P model (see Eq. \eqref{eq:cfl} in the appendix and \cite{lele92} for more details on the time step constraint using compact schemes). Ultimately, the viscoelastic computations require approximately 10 times more CPU resources than the Newtonian computations for a given computational time period.
\section{Viscoelastic turbulence statistics}
\label{sec:viscoelasticstats}
\subsection{Polymer drag reduction}
Since the computations are performed with a constant flow rate by adjusting the pressure gradient, DR is manifested via a decrease in skin friction, i.e. lower $\text{Re}_\tau = \delta/\delta_\nu$ values as drag reduces. Here, we define percentage drag reduction as a negative quantity
\begin{align}
 \label{eq:dragreduction}
 \text{DR} &\equiv \frac{ -\frac{\dd\avg p}{\dd x} - \lrbig{-\frac{\dd\avg p}{\dd x}}\big |_0 } { -\frac{\dd\avg p}{\dd x}\big |_0 } \times 100\%
        = \frac{u_\tau^2 - u_\tau^2\big|_0}{u_\tau^2 \big|_0 } \times 100\% \nonumber\\
       &= \lrbig{\lrbig{\frac{\text{Re}_\tau}{\text{Re}_{\tau0}}}^2 - 1} \times 100\%
\end{align}
with $u_\tau^2 = -\frac{\delta}{\rho}\frac{\dd\avg p}{\dd x}$. Variables with and without subscript 0 in Eq. \eqref{eq:dragreduction} refer to Newtonian\footnote{Any departure from the Newtonian behaviour, i.e. $\sigma_{ij} \propto S_{ij}$, with some constant of proportionality independent of the rate of strain, could be called non-Newtonian.} and viscoelastic fluid flow, respectively. Note that for a direct comparison between the various cases with different skin friction we choose our plots to be presented in terms of $y/\delta$ instead of $y/\delta_\nu$ because $\delta$ is the same for all our computations, whereas $\delta_\nu$ increases with drag reduction.

\begin{figure}[!ht]
  \includegraphics[width=8.5cm]{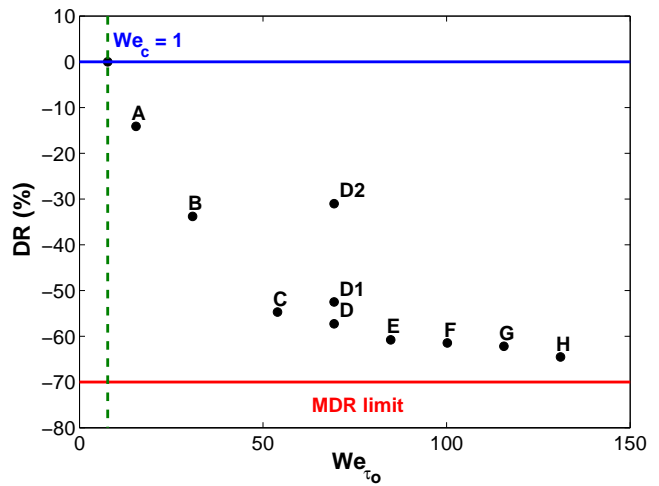}
  \caption{(Colour online) Variation of percentage drag reduction with Weissenberg number.}
  \label{fig:dragreduction}
\end{figure}

Figure \ref{fig:dragreduction} depicts the capability of the current numerical scheme for the FENE-P model to enable stronger polymer-turbulence interactions than artificial diffusion methods. Higher values of percentage drag reduction as function of Weissenberg number are obtained comparing with earlier DNS studies (see Fig. 1b in \cite{minetal03b}) without the need for low $\beta$ values \cite{ptasinskietal03}.
These DR values extend throughout the drag reduction regimes. The MDR limit is approached in this case at $|\text{DR}| \simeq 65\%$ because of the moderate $\text{Re}_c$ in our computations. Even so, this amount of drag reduction falls within the MDR regime, based on the classification of drag reduction by Warholic \etal \cite{warholicetal99}, allowing to study the MDR dynamics of the polymer molecules and their effects on the flow in this asymptotic state.

\subsection{Effects of polymer extensibility and Reynolds number}
\begin{figure}[!ht]
 \includegraphics[width=8.5cm]{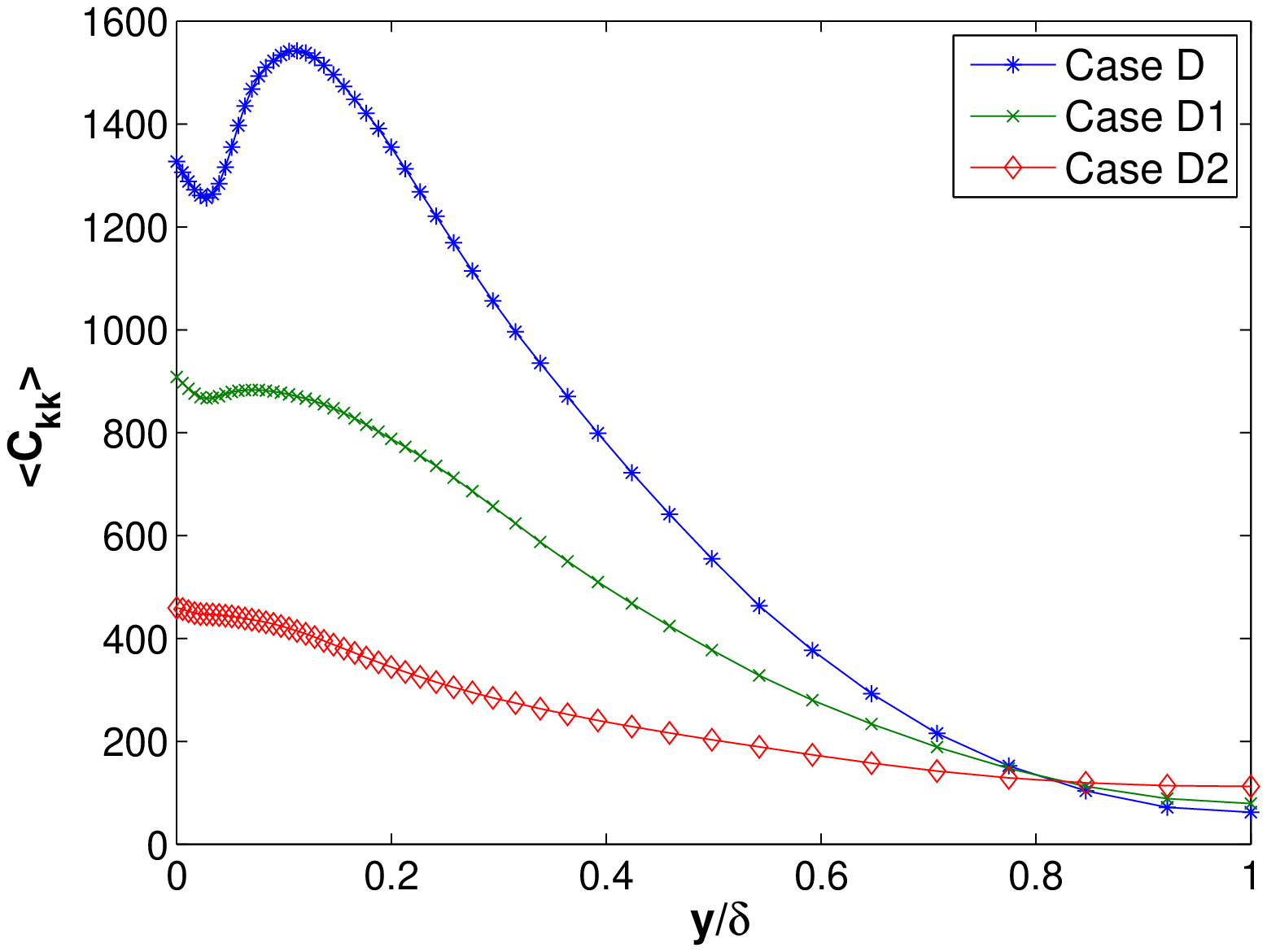}
 \includegraphics[width=8.5cm]{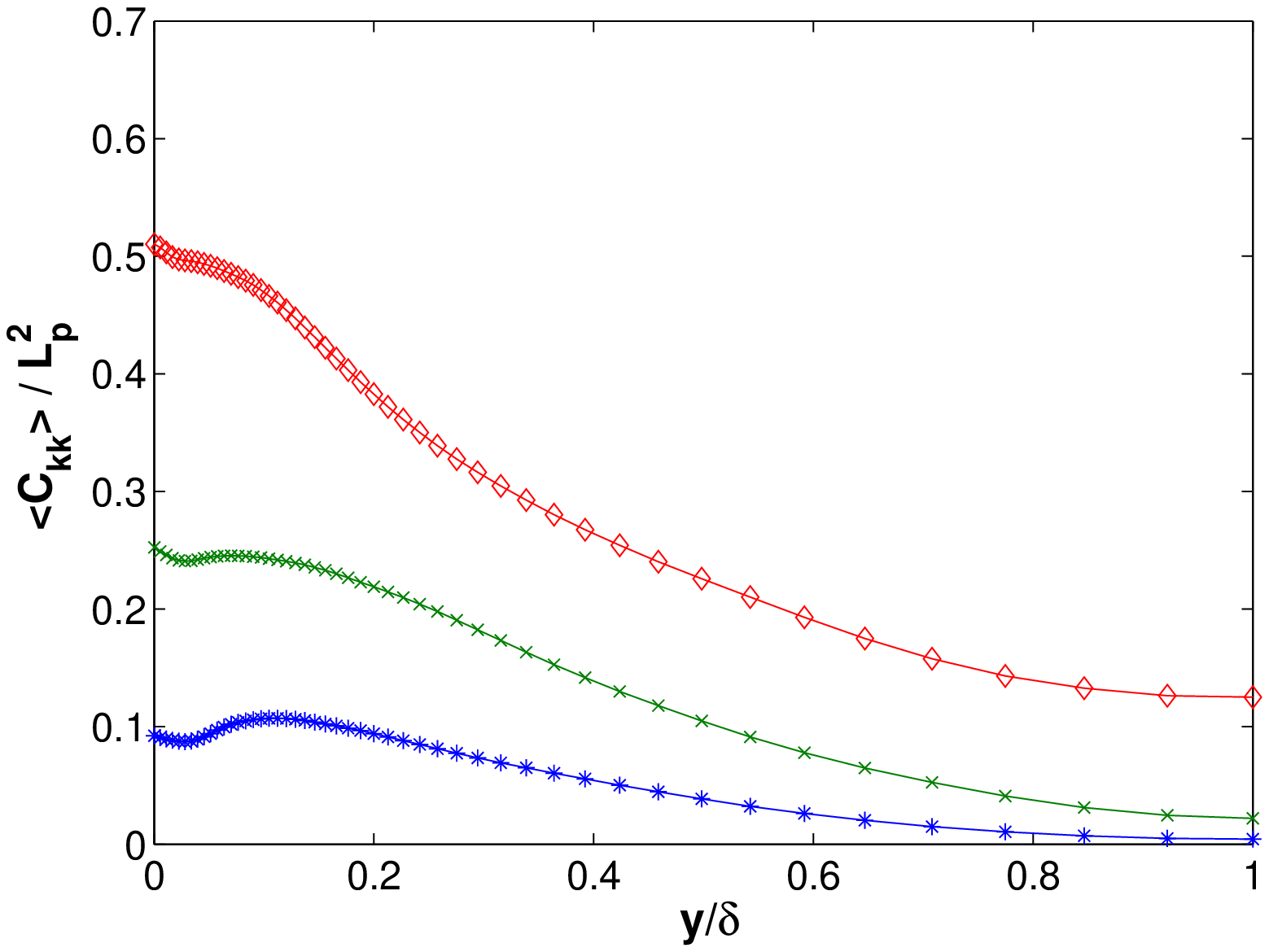}
 \caption{(Colour online) Effect of maximum dumbbell length. Plots of (a) average actual dumbbell extensibility $\avg{C_{kk}}$ and (b) percentage average dumbbell extensibility $\avg{C_{kk}}/L_p^2$ as functions of $y/\delta$. Note: case D ($L_p = 120)$; case D1 ($L_p = 60$); case D2 ($L_p = 30$).}
 \label{fig:Lpeffect}
\end{figure}

The effects of maximum dumbbell extensibility is briefly considered for three different extensibilities $L_p = 30, 60$ and $120$ but the same $\text{We}_c$ and $\text{Re}_c$ (see D cases in Table \ref{tbl:dnspparameters}). Figure \ref{fig:dragreduction} shows that the extent of drag reduction is amplified by longer polymer chains consistent with other DNS studies \cite{dimitropoulosetal98,lietal06}. This effect is related to the fact that the average actual length of the dumbbells, represented by the trace of the conformation tensor $\avg{C_{kk}}$, increases further for larger $L_p$ according to Fig. \ref{fig:Lpeffect}a, inducing stronger influence of the polymers on the flow.
The percentage increase, however, of the polymers extension is less for larger FENE-P dumbbells (see Fig. \ref{fig:Lpeffect}b), suggesting that large polymer molecules could be less susceptible to chain scission degradation, which causes loss of drag reduction in experiments \cite{whitemunghal08}. The near-wall turbulence dynamics play an important role for all three cases, as most of the stretching happens near the wall, where the highest fluctuating strain rates are expected. Eventually, the largest maximum length, i.e. $L_p = 120$, was used for the rest of the computations considered in this work in order to explore the polymer dynamics at effective drag reductions, which are interesting not only fundamentally but also in many real life applications.

\begin{figure}[!ht]
  \includegraphics[width=8.5cm]{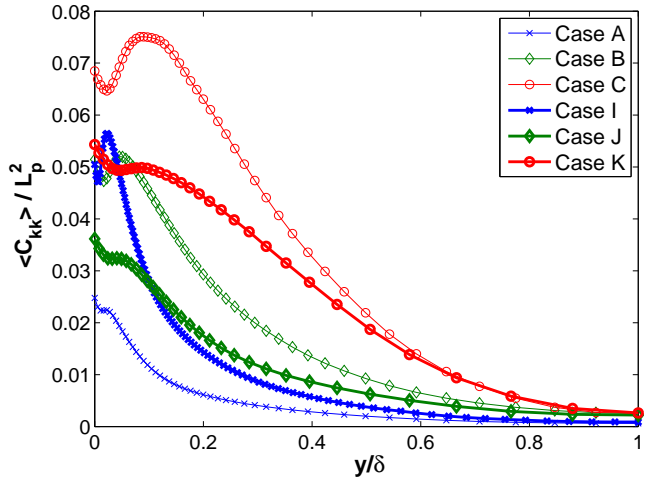}
  \caption{(Colour online) Effect of Reynolds number on percentage average dumbbell extensibility as function of $y/\delta$. Identical symbols correspond to cases with the same $\text{We}_c$ values. Note: Compare case A ($\text{We}_c = 2$, $\text{Re}_c = 4250$) with case I ($\text{We}_c = 2$, $\text{Re}_c = 10400$); case B ($\text{We}_c = 4$, $\text{Re}_c = 4250$) with case J ($\text{We}_c = 4$, $\text{Re}_c = 2750$); case C ($\text{We}_c = 7$, $\text{Re}_c = 4250$) with case K ($\text{We}_c = 7$, $\text{Re}_c = 2750$).}
  \label{fig:Reeffect}
\end{figure}

Based on DNS with artificial diffusion methodology, Housiadas and Beris \cite{housiadasberis03} claim that the extent of drag reduction is rather insensitive to Reynolds numbers in the range between $125 \leq \text{Re}_{\tau_0} \leq 590$ for LDR flows. On the other hand, avoiding the use of artificial diffusion in our study, the Reynolds number dependence on drag reduction for cases with identical $\text{We}_c$ values but different Reynolds numbers, i.e. $\text{Re}_c = 2750, 4250$ and $10400$, is obvious by comparing DR of case A with I and case B with J (LDR regime), as well as case C with K (HDR regime), where the percentage DR increases for higher $\text{Re}_c$ at all instances (see Table \ref{tbl:dnspparameters}). This Reynolds number dependence is further depicted in the polymer dynamics through the profiles of $\avg{C_{kk}}/L_p^2$, which amplify closer to the wall due to more intense strain rates in this region at higher $\text{Re}_c$ and collapse towards the centre of the channel (see Fig. \ref{fig:Reeffect}).
The disparate behaviour of $\avg{C_{kk}}$ with respect to $y/\delta$ due to the Reynolds number dependence is anticipated by the broader spectra of flow time scales that are encountered at higher $\text{Re}_c$ by the dumbbells with fixed relaxation time scale. The fact that the current DNS could capture the Reynolds number dependence on drag reduction and polymer dynamics, emphasises once more the strong polymer-turbulence interactions that can be captured by the present numerical approach even at low levels of drag reduction.

It is essential to note at this point that the intermediate dynamics between the von K\'arm\'an and the MDR law, i.e. the LDR and HDR regimes, are non-universal because they depend on polymer concentration, chemical characteristics of polymers, Reynolds number, etc. \cite{virk75,procacciaetal08}. Here, this is illustrated by the maximum dumbbell length and Reynolds number dependencies of the polymer dynamics in Figs. \ref{fig:Lpeffect} and \ref{fig:Reeffect}, respectively. However, at the MDR limit, which is achieved at $\text{We}_c \gg 1$ and $\text{Re}_c \gg 1$, the dynamics are known to be universal \cite{virk75,procacciaetal08}, i.e. independent of polymer and flow conditions.

\subsection{Mean and fluctuating velocity statistics}
\label{sec:velstats}
\begin{figure}[!ht]
 \includegraphics[width=8.5cm]{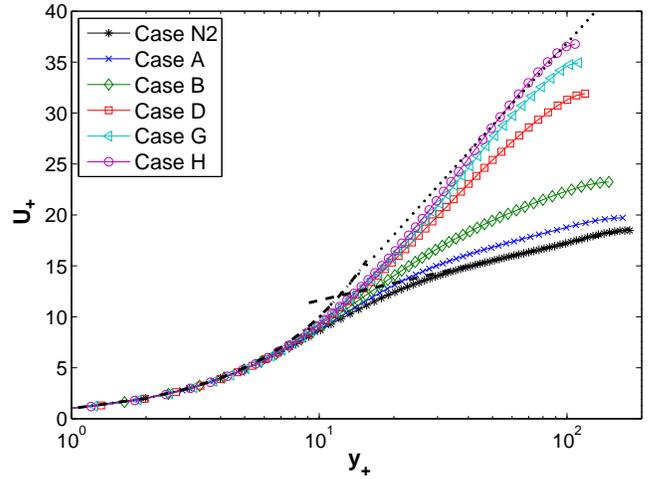}
 \caption{(Colour online) Mean velocity profiles versus $y_+$ for the LDR, HDR and MDR regimes. --\,$\sdot$\,--: $U_+ = y_+$, - - -: $U_+ = \frac{1}{0.41} \log y_+ + 6.0$, $\dotsi$: $U_+ = \frac{1}{11.7} \log y_+ - 17$. Note: case N2 ($\text{DR}=0\%$); case A ($\text{DR}=-14.2\%$); case B ($\text{DR}=-33.8\%$); case D ($\text{DR}=-57.3\%$);  case G ($\text{DR}=-62.1\%$); case H ($\text{DR}=-64.5\%$).}
  \label{fig:velmean}
\end{figure}

\begin{figure}[!ht]
 \includegraphics[width=8.5cm]{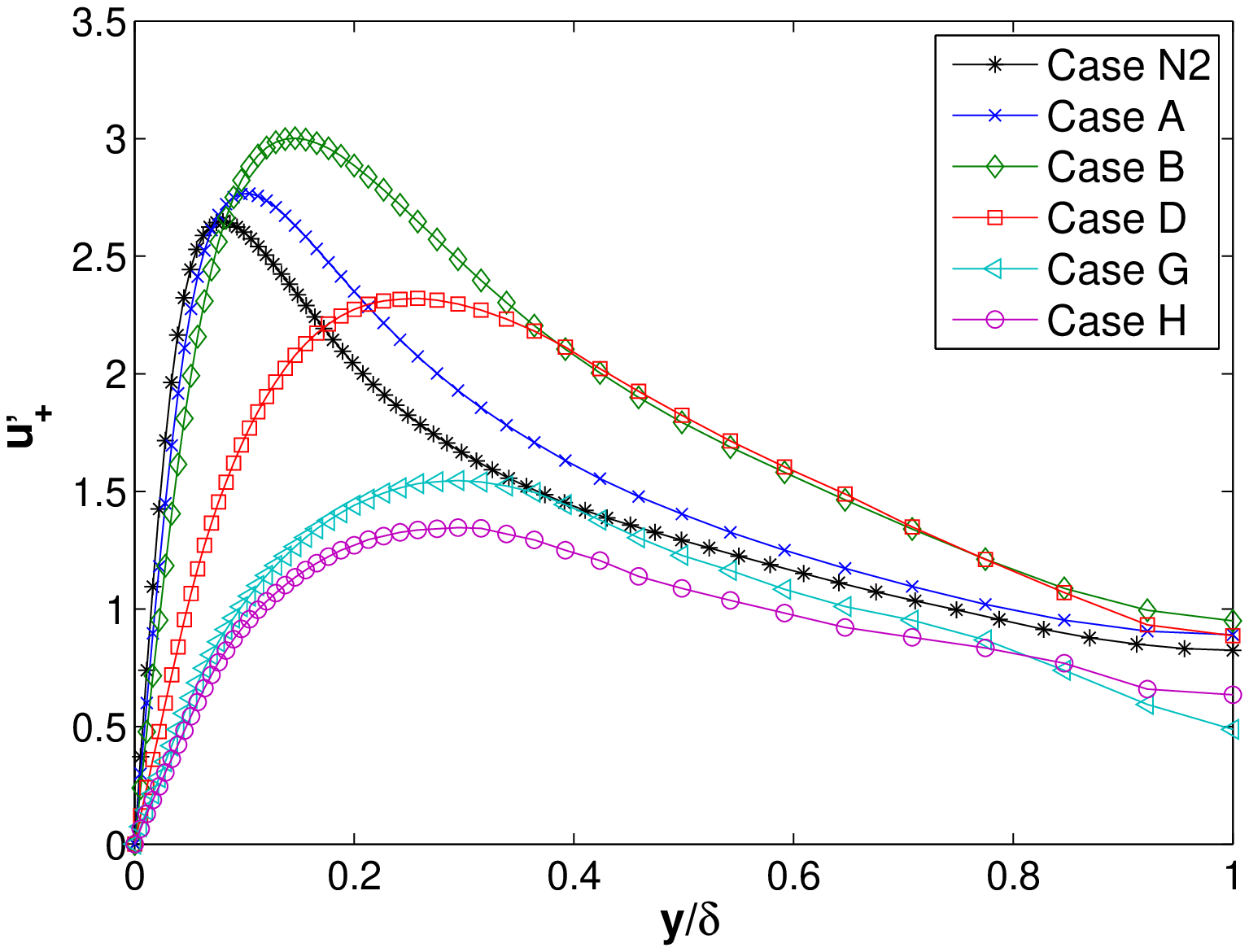}
 \includegraphics[width=8.5cm]{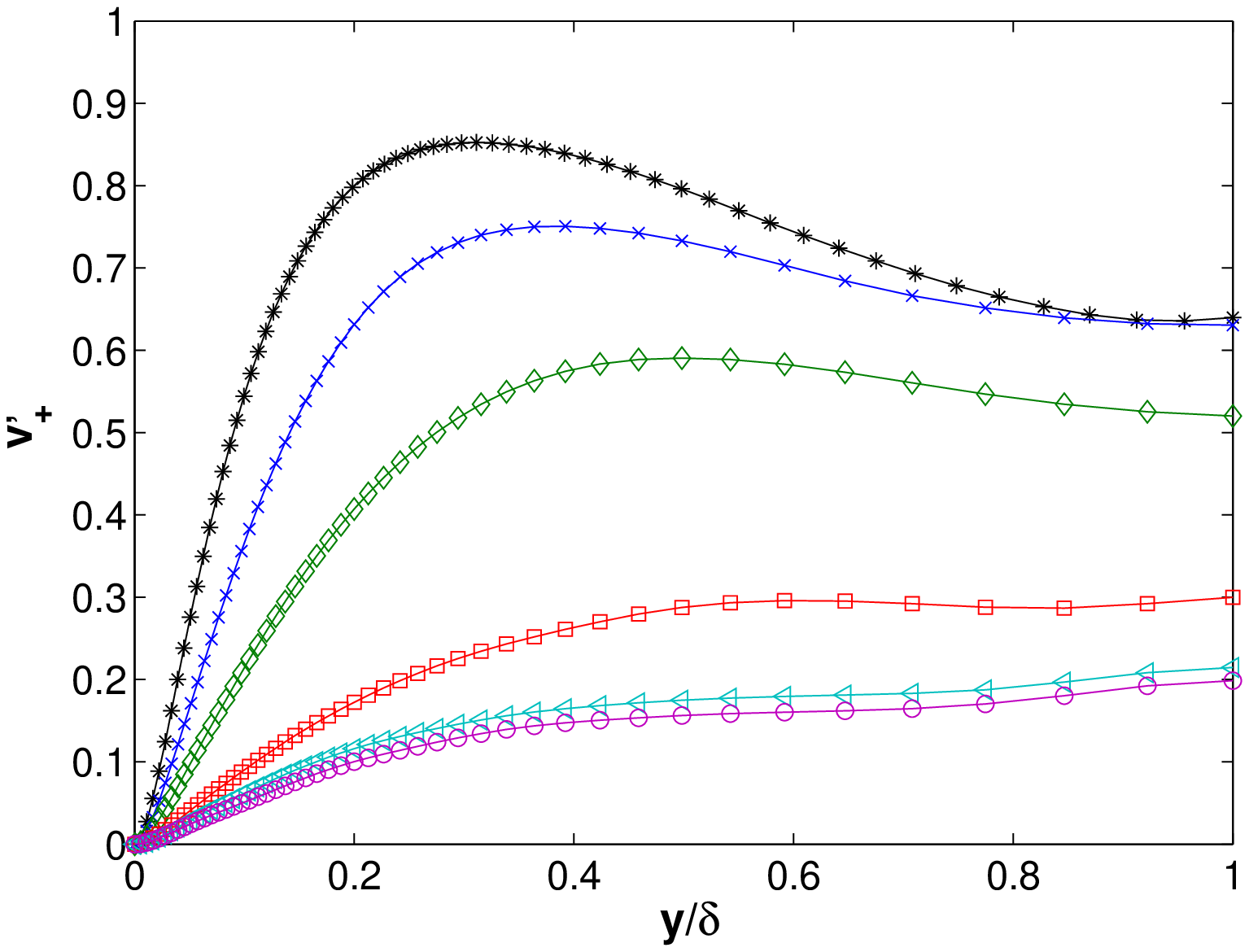}
 \includegraphics[width=8.5cm]{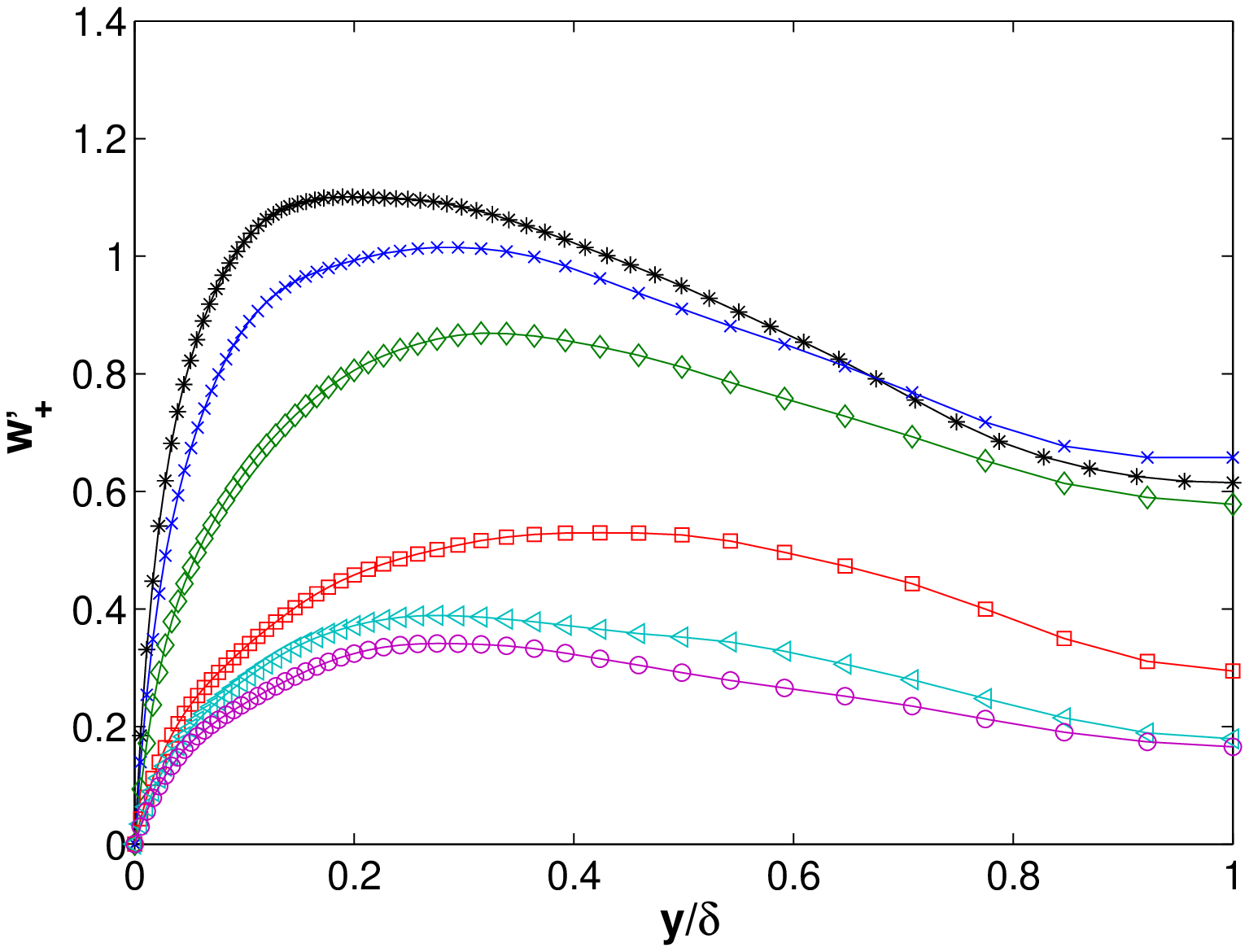}
  \caption{(Colour online) Rms velocity components for the LDR, HDR and MDR regimes. (a) Streamwise $u'_+$, (b) wall-normal $v'_+$ and (c) spanwise $w'_+$ profiles versus $y/\delta$. Note: case N2 ($\text{DR}=0\%$); case A ($\text{DR}=-14.2\%$); case B ($\text{DR}=-33.8\%$); case D ($\text{DR}=-57.3\%$);  case G ($\text{DR}=-62.1\%$); case H ($\text{DR}=-64.5\%$).}
  \label{fig:velrms}
\end{figure}

The picture of drag reduction can be analysed in further detail with the statistics of the turbulent velocity field introduced in Figs. \ref{fig:velmean} and \ref{fig:velrms}. The distinct differences in the statistical trends of the turbulent velocity field between the LDR and HDR regime, that have been observed experimentally, are clearly identified in these results. For clarity, a few indicative cases from the data of Table \ref{tbl:dnspparameters} have been chosen for plotting, representing the LDR, HDR and MDR regimes for different Weissenberg numbers at $\text{Re}_c = 4250$.

According to Fig. \ref{fig:velmean} and noting that $\beta = 0.9$ for all viscoelastic cases, all mean velocity profiles collapse in the viscous sublayer $y_+ \lesssim 10$ to the linear variation $U_+ = \beta^{-1}y_+$, which can be deduced for viscoelastic flows, by rewriting Eq. \eqref{eq:mombalancep} in viscous scales
\begin{equation}
 \label{eq:mombalancep2}
 \beta\frac{\dd U_+}{\dd y_+} - \frac{\avg{u'v'}}{u_\tau^2} + \frac{\avg{\sigma_{12}}}{u_\tau^2} = 1 - \frac{y_+}{\text{Re}_\tau}
\end{equation}
and neglecting the normalised Reynolds and mean polymer shear stress in the viscous sublayer $y_+ \to 0$ (see also section \ref{sec:stressbalance}). Figure \ref{fig:velmean} presents the clear impact of percentage DR on the mean flow with the skin friction decreasing and the mean velocity increasing away from the wall in comparison to the Newtonian case N2 as a result of higher $\text{We}_c$ values at the same $\text{Re}_c$. 
The profile of the Newtonian case N2 is in agreement with the von K\'arm\'an law Eq. \eqref{eq:loglaw}, which does not hold for viscoelastic turbulent flows. Specifically, the curves of cases A and B (LDR regime) are shifted upwards with higher values of the intercept constant $B$, i.e. parallel to the profile of the Newtonian flow (see Fig. \ref{fig:velmean}), increasing DR. This picture is consistent with the phenomenological description by Lumley \cite{lumley69,lumley73}, where the upward shift of the inertial sublayer can be interpreted as a thickening of the buffer or elastic layer for viscoelastic flows, which is equivalent to drag reduction. HDR cases D and G exhibit different statistical behaviour than LDR flows with the slope of the log-region increasing until the MDR asymptote is reached by case H. Overall, the same behaviour across the extent of drag reduction in viscoelastic turbulent flows have been seen in several experimental and numerical results \cite{whitemunghal08}.

Different statistical trends between low and high drag reduction have also been observed experimentally \cite{warholicetal99,ptasinskietal01} for the rms streamwise velocity fluctuations normalised with $u_\tau$. Figure \ref{fig:velrms}a illustrates the growth of the peak in the profile of $u'_+$ for LDR case A and B at low $\text{We}_c$ and a notable decrease for the rest of the cases at HDR/MDR with high $\text{We}_c$ values. The peaks move monotonically away from the wall throughout the drag reduction regimes indicating the thickening of the elastic layer, which is compatible with the behaviour of the mean velocity profile. 

Note that this is the first time that a DNS computation can so distinctly attain this behaviour. This is attributed to the accurate shock-capturing numerical scheme we applied for the FENE-P model in this study. It has to be mentioned however that there have been three earlier studies \cite{minetal03b,ptasinskietal03,dubiefetal04}, which use the artificial diffusion algorithms for FENE-P and showed similar but not as clear trends for $u'_+$ in a DNS of viscoelastic turbulent channel flow. In fact, Min \etal \cite{minetal03b} reached the HDR/MDR regime at roughly $|\text{DR}| \simeq 40\%$, clearly very low to afford the correct dynamics and Ptasinski \etal \cite{ptasinskietal03} had to use $\beta = 0.4$ to approach HDR/MDR, encountering considerable shear-thinning effects. It is interesting to mention that other recent studies \cite{handleretal06,lietal06}, using the artificial diffusion methodology, with more extensive Weissenberg number data and high $\beta$ values, have not been able to obtain this transition effect on the statistics of $u'_+$ between the drag reduction regimes.

Finally, the wall-normal $v'_+$ and spanwise $w'_+$ rms velocity fluctuations in Figs. \ref{fig:velrms}b and \ref{fig:velrms}c, respectively, are continuously attenuated while DR is enhanced by increasing the polymer relaxation time scale. Again, the monotonic displacement of their peaks towards the centre of the channel as drag reduction amplifies is consistent with that of the mean velocity profile and with experimental and other numerical studies \cite{whitemunghal08}.

\subsection{Fluctuating vorticity statistics}
\label{sec:vortstats}
\begin{figure}[!ht]
   \includegraphics[width=8.5cm]{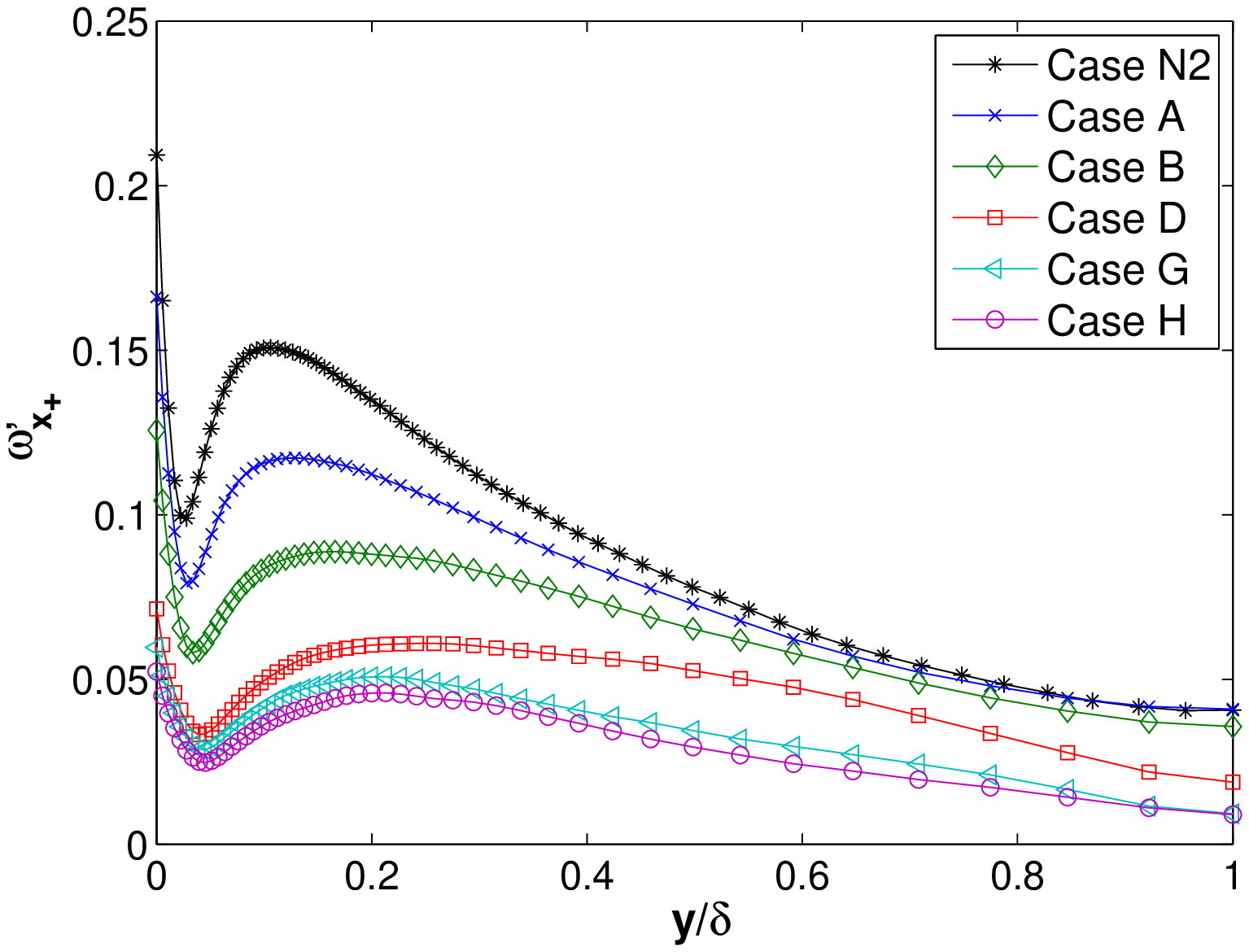}
   \includegraphics[width=8.5cm]{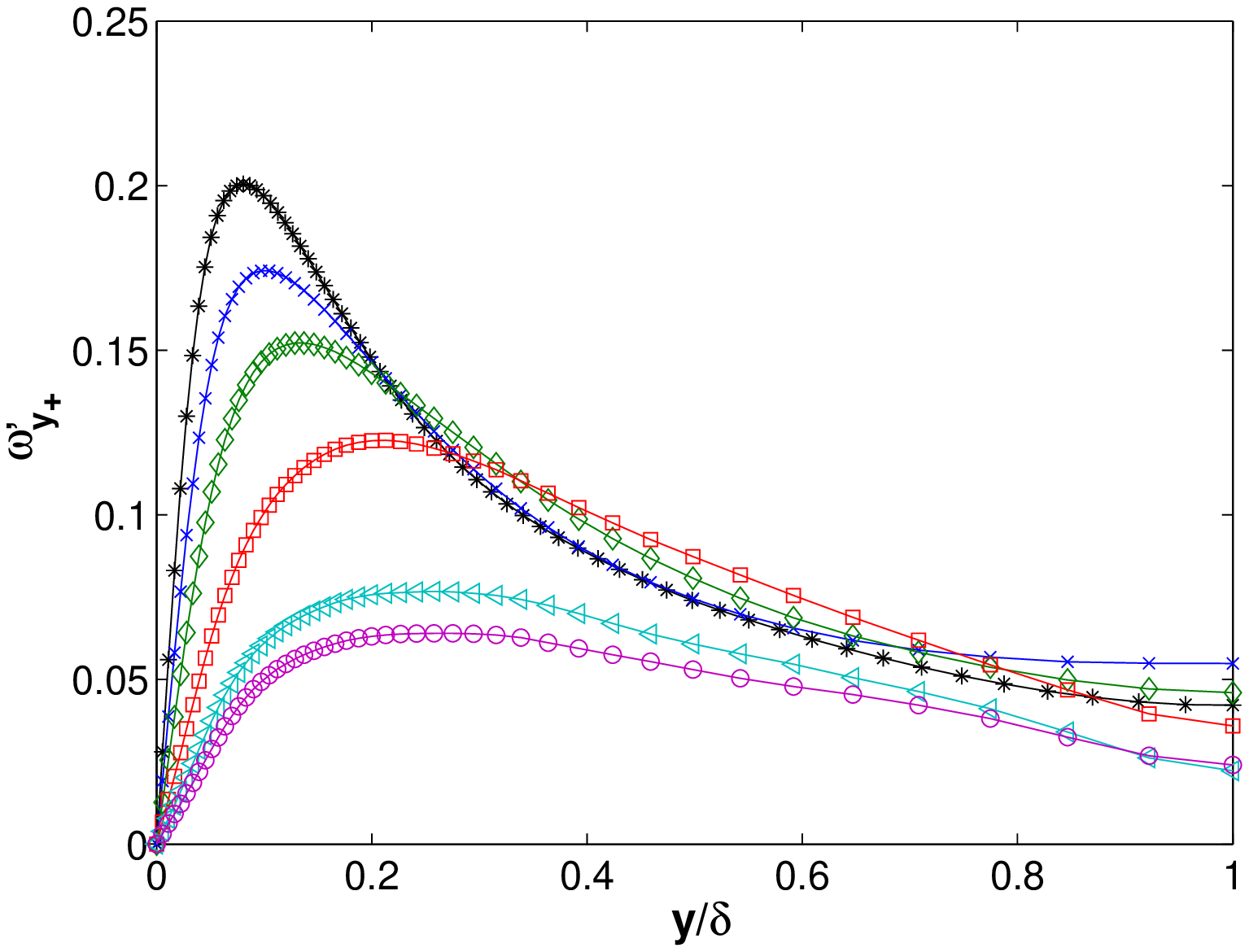}
   \includegraphics[width=8.5cm]{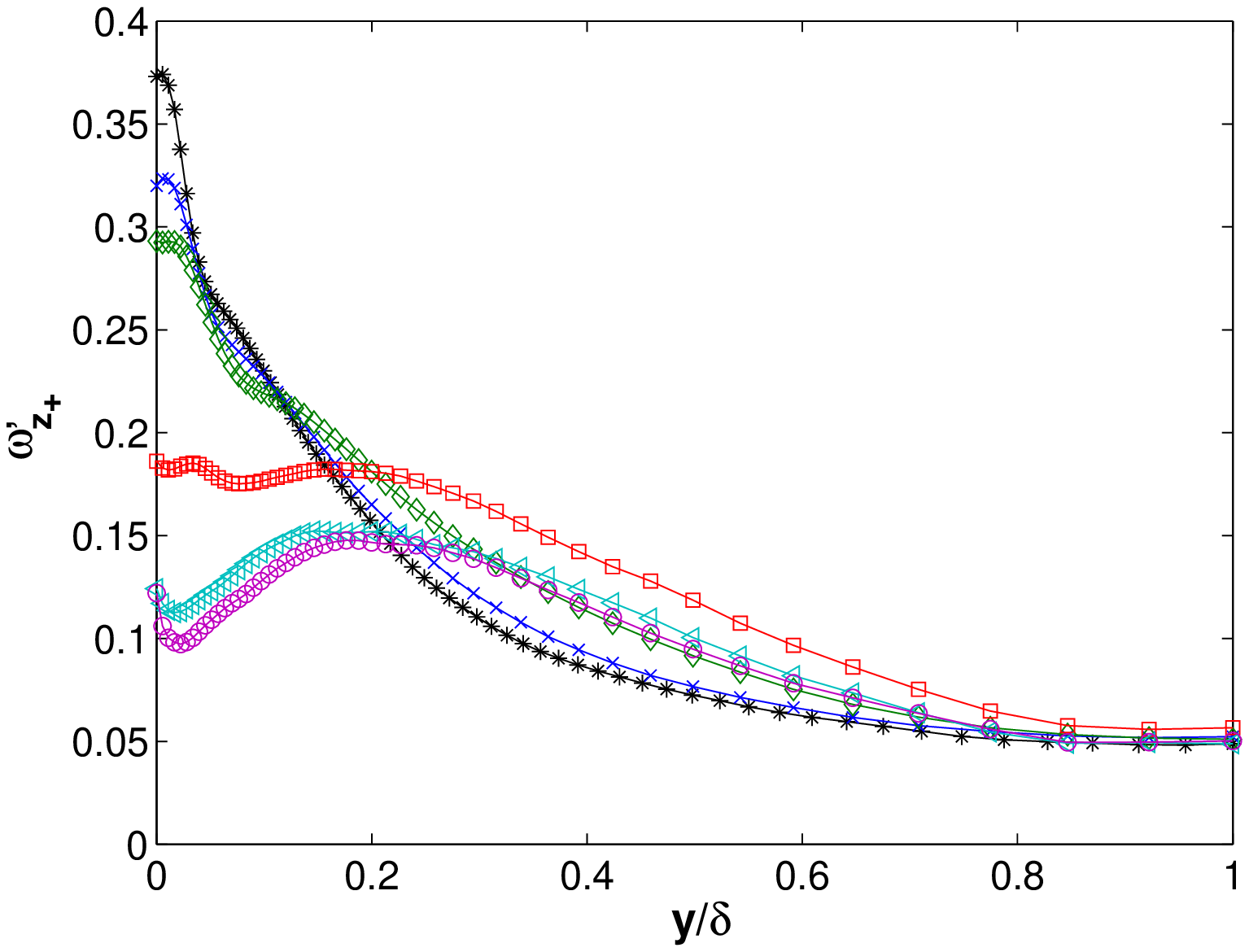}
  \caption{(Colour online) Rms vorticity profiles for the LDR, HDR and MDR regimes. (a) Streamwise $\omega'_{x_+}$, (b) wall-normal $\omega'_{y_+}$ and (c) spanwise $\omega'_{z_+}$ profiles versus $y/\delta$. Note: case N2 ($\text{DR}=0\%$); case A ($\text{DR}=-14.2\%$); case B ($\text{DR}=-33.8\%$); case D ($\text{DR}=-57.3\%$);  case G ($\text{DR}=-62.1\%$); case H ($\text{DR}=-64.5\%$).}
  \label{fig:vortrms}
\end{figure}

The rms statistics of the fluctuating vorticity field normalised by viscous scales, i.e. $\bm \omega'_+ \equiv \bm \omega'\delta_\nu / u_\tau$, are presented in Fig. \ref{fig:vortrms} for representative cases from Table \ref{tbl:dnspparameters} at various levels of drag reduction.
The streamwise vorticity fluctuations $\omega'_{x_+}$ demonstrate a persistent attenuation along the normalised distance $y/\delta$ as drag reduction enhances due to the increase of $\text{We}_c$ (see Fig. \ref{fig:vortrms}a). In the near-wall region $y/\delta < 0.2$ of Fig. \ref{fig:vortrms}a there is a characteristic local minimum and maximum that could be interpreted as corresponding to the average edge and centre of the streamwise vortices, respectively \cite{kmm87,lietal06}. Then, the average size of these large streamwise vortices is roughly equal to the distance between these two peaks. The fact that these peaks are displaced away from each other and at the same time away from the wall, as DR builds up, implies an increase in the average size of the streamwise vortices and a thickening of the buffer layer, respectively, in agreement with earlier works \cite{sureshkumaretal97,lietal06,kimetal07,whitemunghal08}. The attenuation in the intensity of $\omega'_{x_+}$ provides evidence for a drag reduction mechanism based on the suppression of the near-wall counter-rotating steamwise vortices \cite{kimetal07,kimetal08}, which underpin considerable amount of the turbulence production \cite{kimetal71}.

The wall-normal rms vorticity is zero at the wall due to the no-slip boundary condition and reaches its peak within the buffer layer (see Fig. \ref{fig:vortrms}b). The intensity of $\omega'_{y_+}$ is reduced for all levels of drag reduction according to Fig. \ref{fig:vortrms}b, with the position of the near-wall peaks moving towards the centre of the channel as $\text{We}_c$ becomes larger, representing once more the thickening of the elastic layer in a consistent way. Most of the inhibition of $\omega'_{y_+}$ happens near the wall and slightly towards the centre of the channel only for the HDR/MDR cases G and H, i.e. for $|\text{DR}| > 60\%$. 

Figure \ref{fig:vortrms}c shows a more interesting behaviour for  $\omega'_{z_+}$, where the spanwise vorticity fluctuations decrease in the near-wall region $y/\delta \lesssim 0.2$ and increase further away while drag reduces. This effect may be related to the transitional behaviour of $u'_+$ between the LDR and HDR/MDR regimes (see Fig. \ref{fig:velrms}a) plus the continuous drop of $v'_+$ (see Fig. \ref{fig:velrms}b) in viscoelastic drag reduced flows. As a final note, $\omega'_{z_+} > \omega'_{x_+} > \omega'_{y_+}$ in the viscous sublayer, i.e. $y/\delta < 0.05$ for all cases and $\omega'_{z_+} \simeq \omega'_{x_+} \simeq \omega'_{y_+}$ in the inertial and outer layer for the Newtonian case N2. However, $\omega'_{z_+} > \omega'_{y_+} > \omega'_{x_+}$ away from the wall when drag reduces for viscoelastic flows, which manifests the dominance of small scale anisotropy in the inertial and outer layer at HDR and MDR.

\section{Conformation and polymer stress tensor}
\label{sec:conftensor}
\begin{figure}[!ht]
 \includegraphics[width=8.5cm]{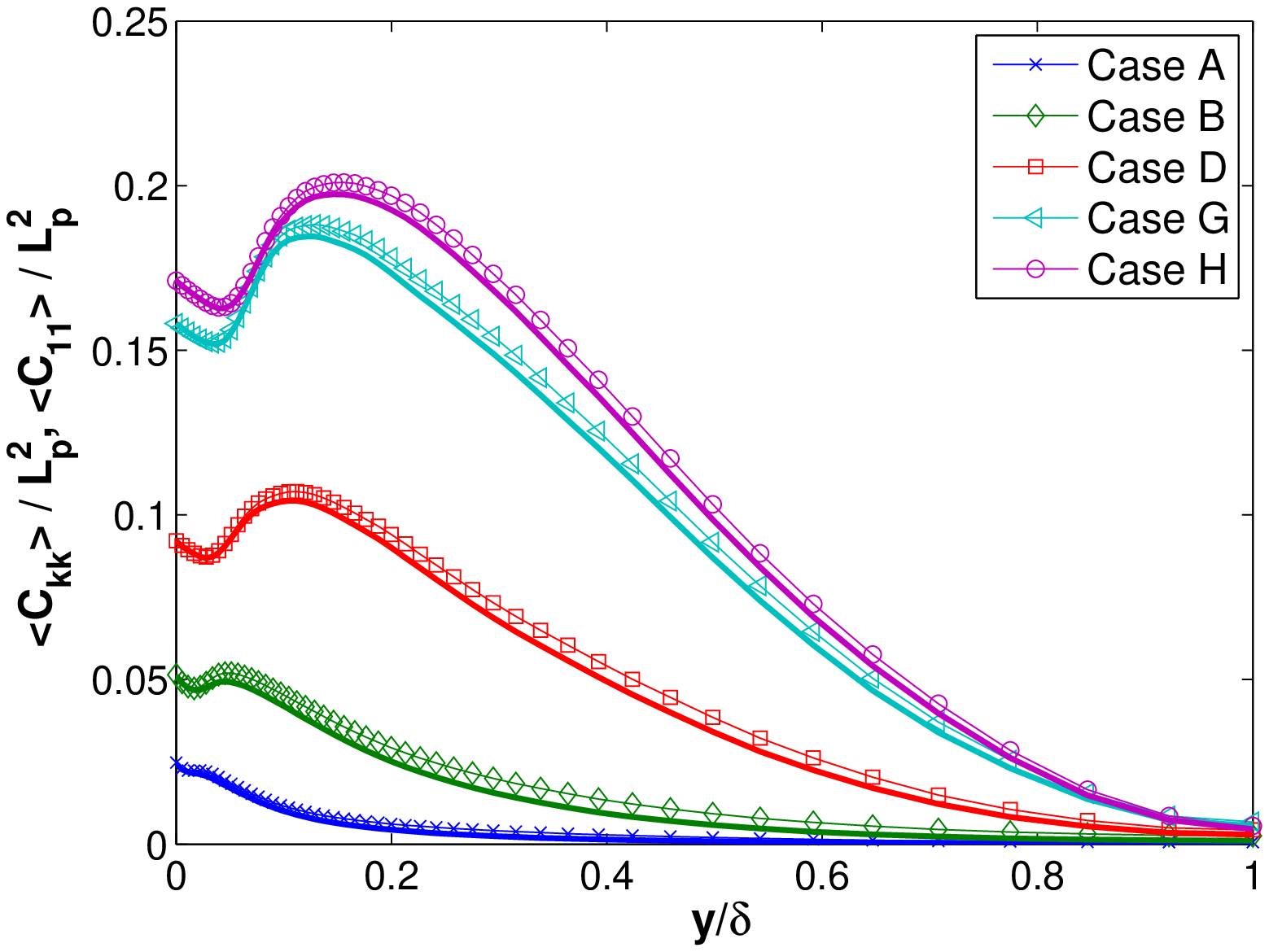}
 \includegraphics[width=8.5cm]{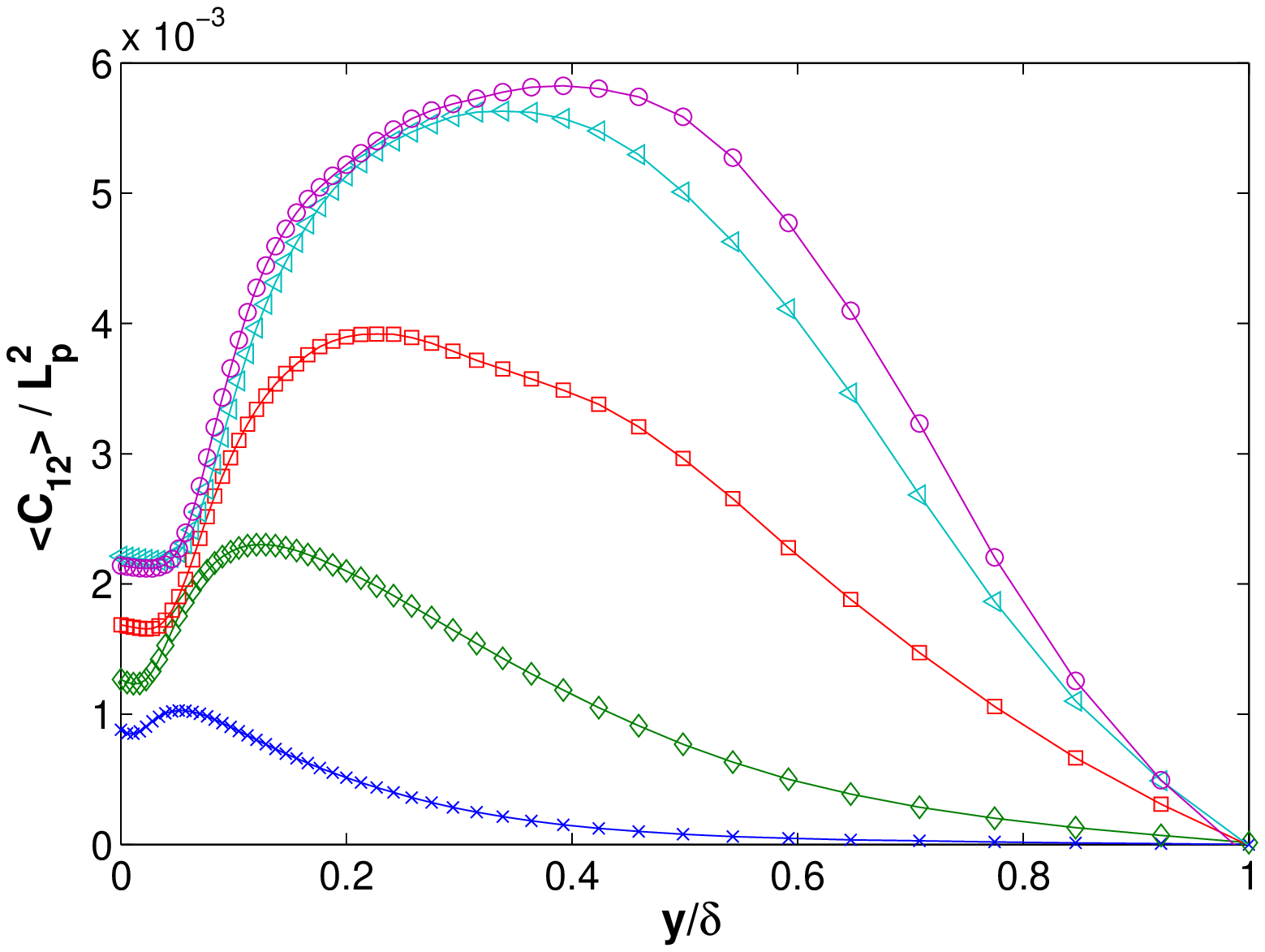}
 \caption{(Colour online) Profiles of (a) $\avg{C_{kk}}/L_p^2$ (line-symbols) and $\avg{C_{11}}/L_p^2$ (solid lines), (b) $\avg{C_{12}}/L_p^2$ as functions of $y/\delta$ for the LDR, HDR and MDR regimes. Note: case A ($\text{DR}=-14.2\%$); case B ($\text{DR}=-33.8\%$); case D ($\text{DR}=-57.3\%$);  case G ($\text{DR}=-62.1\%$); case H ($\text{DR}=-64.5\%$).}
  \label{fig:confab}
\end{figure}

\begin{figure}[!ht]
 \includegraphics[width=8.5cm]{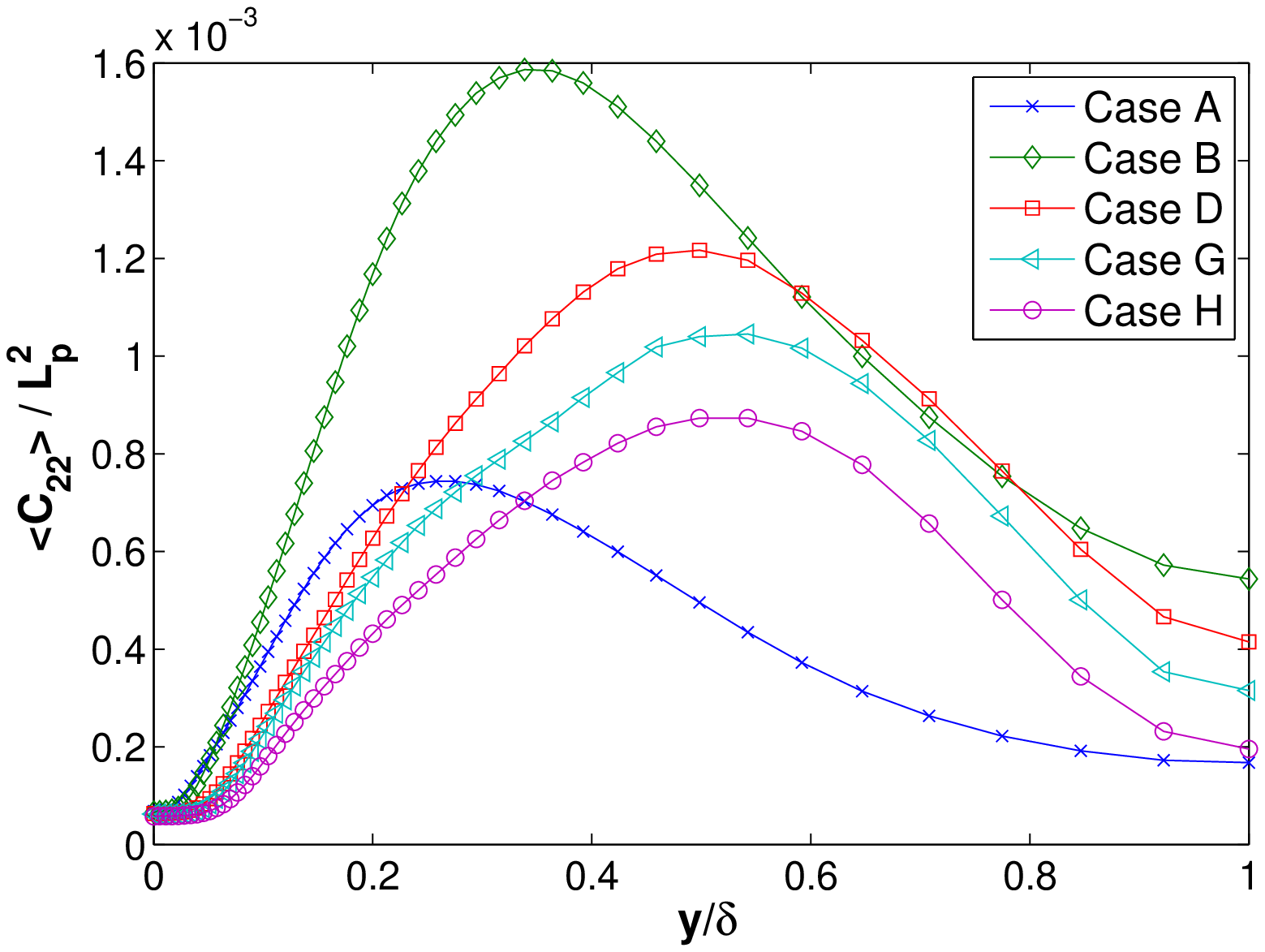}
 \includegraphics[width=8.5cm]{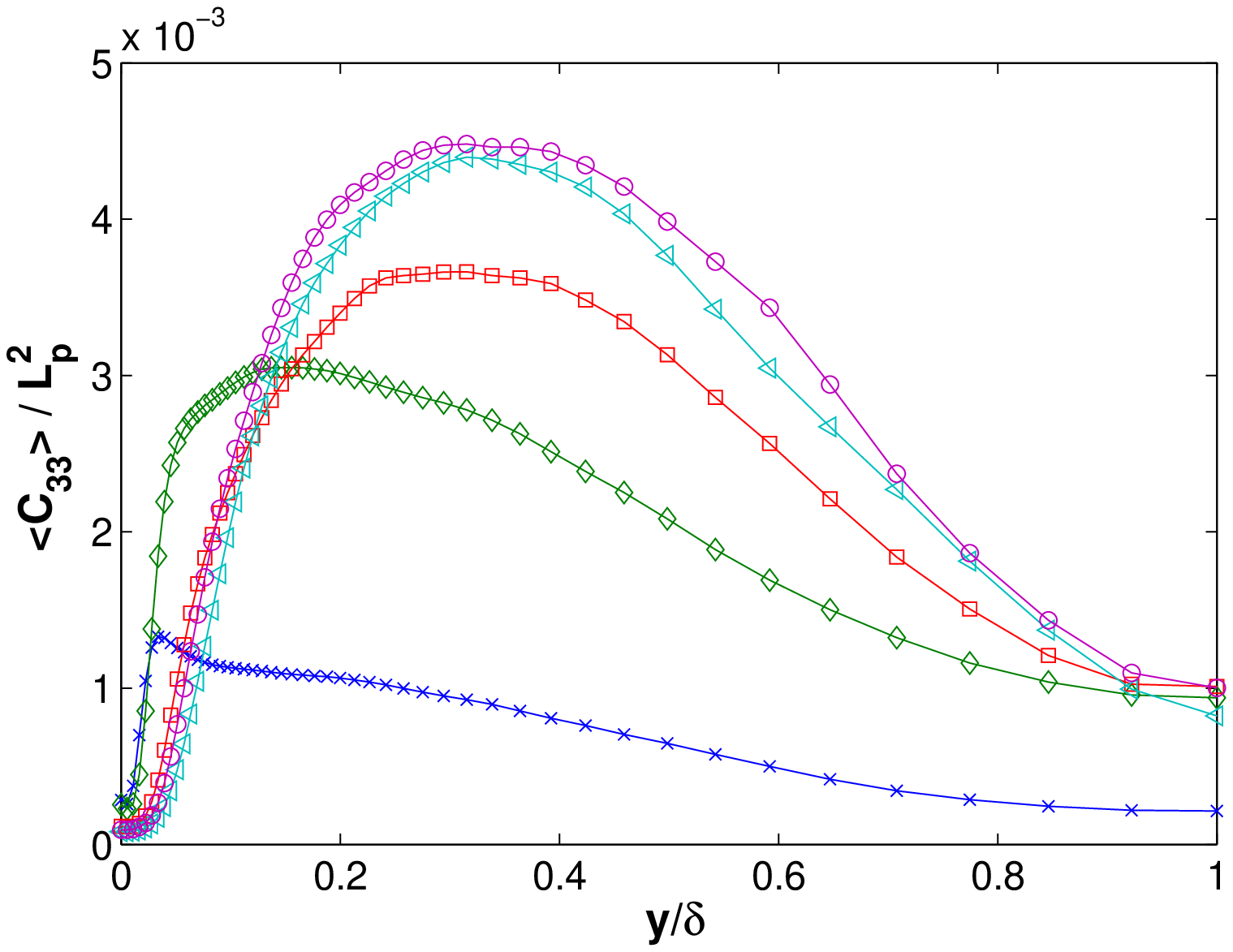}
 \caption{(Colour online) Profiles of (a) $\avg{C_{22}}/L_p^2$ and (b) $\avg{C_{33}}/L_p^2$ as functions of $y/\delta$ for the LDR, HDR and MDR regimes. Note: case A ($\text{DR}=-14.2\%$); case B ($\text{DR}=-33.8\%$); case D ($\text{DR}=-57.3\%$);  case G ($\text{DR}=-62.1\%$); case H ($\text{DR}=-64.5\%$).}
 \label{fig:confcd}
\end{figure}

Before looking at the mean momentum and energy balance, the study of the conformation tensor field is essential to get an understanding of the polymer dynamics in support of the results provided by this new numerical method for the FENE-P model in turbulent channel flow. The symmetries in the flow geometry determine properties of tensor components in the average sense \cite{pope00}. In the current DNS of turbulent channel flow, statistics are independent of the $z$ direction and the flow is also statistically invariant under reflections of the $z$ co-ordinate axis. Therefore, for the probability density function $f(\bm Q;\bm x, t)$ of a vector $\bm Q$, these two conditions imply $\pd f / \pd z = 0$ and $f(Q_1,Q_2,Q_3;x,y,z,t) = f(Q_1,Q_2,-Q_3;x,y,-z,t)$. Then, at $z = 0$ reflectional symmetry suggests that $\avg{Q_3} = -\avg{Q_3} \Rightarrow \avg{Q_3} = 0$ and similarly for $\avg{Q_1Q_3} = \avg{Q_2Q_3} = 0$. So, in this case the mean conformation tensor reduces to
\begin{equation}
 \avg{C_{ij}} = 
  \begin{pmatrix}
   \avg{C_{11}} & \avg{C_{12}} & 0            \\
   \avg{C_{12}} & \avg{C_{22}} & 0            \\
              0 &            0 & \avg{C_{33}}
  \end{pmatrix}
\end{equation}
where the non-zero components scaled with $L_p$ are presented in Figs. \ref{fig:confab} and \ref{fig:confcd} with respect to $y/\delta$ for cases at various drag reduction regimes (see Table \ref{tbl:dnspparameters}). The zero components in our study have been found to be zero within the precision accuracy. Turbulent channel flow is also statistically symmetric about the plane $y = \delta$. Therefore, this reflectional symmetry imposes $f(Q_1,Q_2,Q_3;x,y,z,t) = f(Q_1,-Q_2,Q_3;x,-y,z,t)$, which implies that the normal components of $\avg{C_{ij}}$ are even functions and $\avg{C_{12}}$ is an odd function comparable to the Reynolds stress tensor components.

The normalised trace of the mean conformation tensor $\avg{C_{kk}} / L_p^2$ is plotted in Fig. \ref{fig:confab}a together with $\avg{C_{11}} / L_p^2$. Notice that the dominant contribution in the trace comes from $\avg{C_{11}}$, i.e. $\avg{C_{kk}} \simeq \avg{C_{11}}$ at all Weissenberg numbers, reflecting on average a strong preferential orientation of the stretched dumbbells along the streamwise direction. The fact that $\avg{C_{11}} \gg \avg{C_{12}} \simeq \avg{C_{33}} > \avg{C_{22}}$ denotes the strong anisotropic behaviour of the mean conformation tensor caused by the mean shear in turbulent channel flow. This anisotropy influences the statistics of the fluctuating velocity field particularly at small scales, as was mentioned in section \ref{sec:vortstats}. The curves of $\avg{C_{11}} / L_p^2$ and consequently of $\avg{C_{kk}} / L_p^2$ constantly rise with most of the stretching happening close to the wall and growing towards the centre of the channel, since higher values of polymer time scale are influenced from a wider spectrum of flow time scales. A local minimum and a maximum emerge in the near-wall region $y/\delta < 0.2$, induced by the streamwise vortices \cite{dubiefetal04,dimitropoulosetal05}. These peaks move apart from each other and away from the wall for higher $\text{We}_c$ values. Figure \ref{fig:confab}a also shows that the amplitudes of these peaks seem inversely proportional to the peak amplitudes of $\omega'_{x_+}$ as drag reduces (see also Fig. \ref{fig:vortrms}a).

Moreover, as $\text{We}_c$ increases the profiles of $\avg{C_{12}} / L_p^2$ and $\avg{C_{33}} / L_p^2$ amplify, reaching their peaks at not much different $y/\delta$ for each $\text{We}_c$ case (see Figs. \ref{fig:confab}b and \ref{fig:confcd}b). In particular, the values of $\avg{C_{12}} / L_p^2$ at the wall are dependent on the polymer relaxation time scale unlike for $\avg{C_{33}} / L_p^2$. On the other hand, the values of $\avg{C_{33}} / L_p^2$ depend on Weissenberg number at $y = \delta$ in contrast to $\avg{C_{12}} / L_p^2$, which is zero for all cases because of the symmetry mentioned earlier. The behaviour of $\avg{C_{22}} / L_p^2$ in Fig. \ref{fig:confcd}a is more peculiar with respect to $\text{We}_c$, with the profiles increasing within the LDR regime and attenuate for HDR and MDR cases, in a similar manner to $u'_+$ (see Fig. \ref{fig:velrms}a). Its peak values are achieved closer to the core of the channel in comparison to the rest of the conformation tensor components. This points out the different flow time scales that are important for $\avg{C_{22}}$, exemplifying the complex dynamics of the polymers, even in this simple mechanical model.

It is interesting to mention that the components of $\avg{C_{ij}}$ have different asymptotic rates of convergence towards the limit $\text{We}_c \to \infty$. It is known that for $\text{We}_c \gg 1$ the upper bound for the trace is $\avg{C_{kk}} \leq L_p^2$ and subsequently in our case $\avg{C_{11}} \lesssim L_p^2$ (see Fig. \ref{fig:confab}a), where this upper bound is far from achieved in our computations. This result demonstrates that highly stretched polymers are not required for the manifestation of drag reduction or even of the MDR asymptote, as de Gennes \cite{degennes90} claims against Lumley's \cite{lumley69} assumption of a coil-stretch transition, i.e. highly stretched polymer molecules, for the enhancement of intrinsic viscosity. The components $\avg{C_{12}} / L_p^2$ and $\avg{C_{33}} / L_p^2$ seem to have almost reached their asymptotic limit with the MDR case H according to Figs. \ref{fig:confab}b and \ref{fig:confcd}b, respectively. Finally, $\avg{C_{22}} / L_p^2$ has not yet converged to its limit, decreasing with a slow rate towards very small values for high $\text{We}_c$. In fact, it has been argued theoretically that $\avg{C_{22}} \to 0$ in the limit of infinite Weissenberg number \cite{lvovetal05,procacciaetal08}.

\begin{figure}[!ht]
  \includegraphics[width=8.5cm]{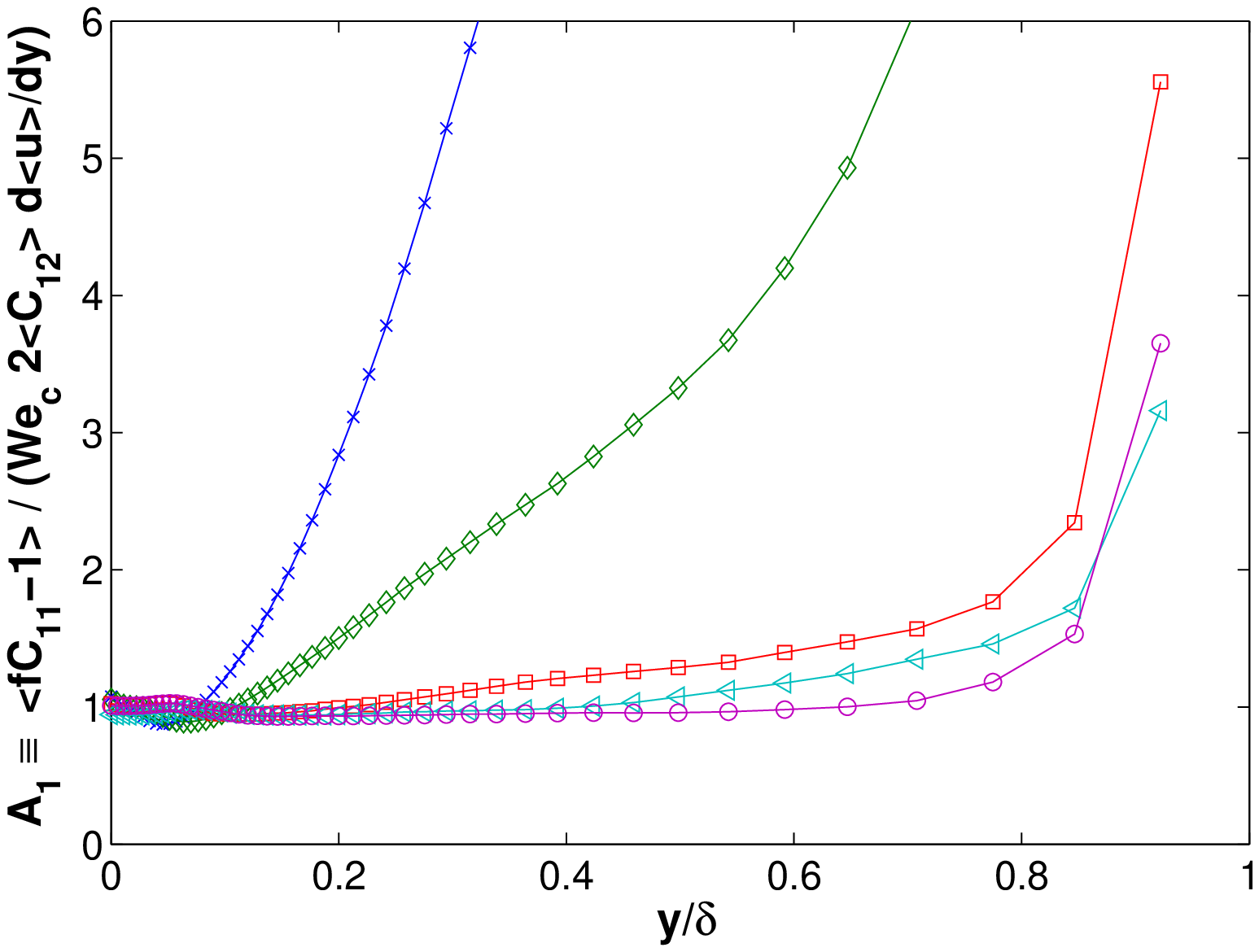}
  \includegraphics[width=8.5cm]{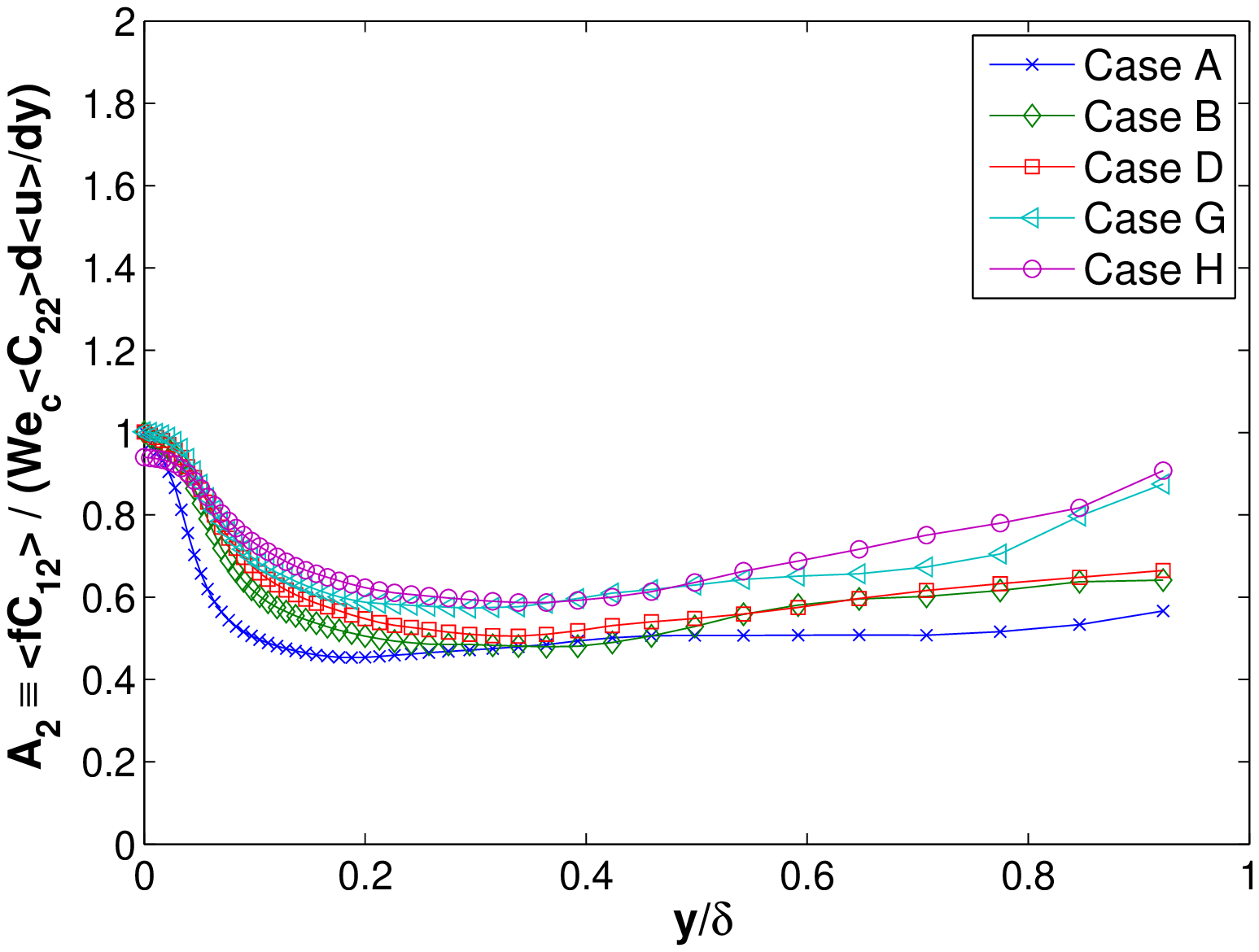}
 \caption{(Colour online) Scalings of the compensated polymer stress components (a) $A_1 \equiv \avg{\sigma_{11}} / \lrbig{\frac{1-\beta}{\text{Re}_c}2\avg{C_{12}}\frac{\dd \avg{u}}{\dd y}}$ and (b) $A_2 \equiv \avg{\sigma_{12}} / \lrbig{\frac{1-\beta}{\text{Re}_c}\avg{C_{22}}\frac{\dd \avg{u}}{\dd y}}$ with respect to $y/\delta$. Note: case A ($\text{DR}=-14.2\%$); case B ($\text{DR}=-33.8\%$); case D ($\text{DR}=-57.3\%$);  case G ($\text{DR}=-62.1\%$); case H ($\text{DR}=-64.5\%$).}
 \label{fig:confscaling}
\end{figure}

Polymer stresses are non-linear with respect to the conformation tensor and their asymptotic scaling with Weissenberg number is a key element for the understanding of the polymer dynamics at MDR. Hence, following Benzi \etal \cite{benzietal06} consider the FENE-P model integrated over the $x$, $z$ spatial directions and time, assuming statistical stationarity and homogeneity in $x$ and $z$
\begin{align}
 \avg{u_2 \pd_{x_2} C_{ij}} &= \avg{C_{ik} \pd_{x_k} u_j} + \avg{C_{jk} \pd_{x_k} u_i} \nonumber\\
 &- \frac{1}{\text{We}_c}\avg{f(C_{kk})C_{ij} - \delta_{ij}}.
\end{align}
Then, taking the Reynolds decomposition of the velocity field $u_i = \avg{u_i} + u'_i$, one obtains
\begin{equation}
 \frac{1}{\text{We}_c}\avg{f(C_{kk})C_{ij} - \delta_{ij}} = \avg{C_{ik}} \pd_{x_k}\avg{u_j} + \avg{C_{jk}} \pd_{x_k}\avg{u_i} + Q_{ij}
\end{equation}
where $Q_{ij} = \avg{C_{ik} \pd_{x_k} u'_j} + \avg{C_{jk} \pd_{x_k} u'_i} - \avg{u'_2 \pd_{x_2} C_{ij}}$. Therefore, the average polymer stress tensor defined by Eqs. \eqref{eq:nondimpolystress} takes the form
\begin{align}
 &\avg{\sigma_{ij}} = \nonumber\\
 & = \frac{1-\beta}{\text{Re}_c}
  \begin{pmatrix}
   2\avg{C_{12}}\frac{\pd\avg{u_1}}{\pd x_2} + Q_{11} &
    \avg{C_{22}}\frac{\pd\avg{u_1}}{\pd x_2} + Q_{12} & Q_{13} \\
    \avg{C_{22}}\frac{\pd\avg{u_1}}{\pd x_2} + Q_{12} & Q_{22} & Q_{23} \\
   Q_{13} & Q_{23} & Q_{33}
  \end{pmatrix}.
\end{align}
Now, the important assumption at the limit of a local Weissenberg number $\text{We}_S \equiv \tau_p \frac{\dd}{\dd y}\avg{u} \to \infty$ is that $Q_{11}$ and $Q_{12}$ can be neglected, considering the polymers to be stiff, i.e. $C_{ij} \to \avg{C_{ij}}$, mostly in the main stretching directions and the correlations between fluctuating conformation tensor and velocity fields in the other Cartesian directions to remain minimal at this limit. In this case, as a result
\begin{align}
 \avg{\sigma_{11}} &= A_1 \frac{1-\beta}{\text{Re}_c} 2\avg{C_{12}}\pd_{x_2}\avg{u_1} \\
 \label{eq:a2}
 \avg{\sigma_{12}} &= A_2 \frac{1-\beta}{\text{Re}_c} \avg{C_{22}}\pd_{x_2}\avg{u_1}
\end{align}
where $A_1$ and $A_2$ are expected to be independent of $y$ and equal to 1 at some intermediate region in the flow as $\text{We}_S \gg 1$. This hypothesis is checked in Fig. \ref{fig:confscaling} against various viscoelastic DNS from Table \ref{tbl:dnspparameters}.

Figure \ref{fig:confscaling}a shows clearly that $A_1$ tends to a constant and reaches 1 in the region $0 \lesssim y/\delta \lesssim 0.8$ for high $\text{We}_S$ values justifying that $Q_{11}$ can be neglected for HDR and MDR cases. Note that $A_1$ deviates from 1 towards the centre of the channel because $\text{We}_S$ becomes small in this region. $A_2$ is approximately independent of $y$ in some intermediate region for almost all cases and appears to tend towards 1 as $\text{We}_S$ increases (see Fig. \ref{fig:confscaling}b). However, the polymer relaxation time scales used in this study are not sufficiently high for $A_2 \to 1$. So, for our DNS results the polymer shear stress can be considered to be $\avg{\sigma_{12}} \propto \frac{1-\beta}{\text{Re}_c} \avg{C_{22}}\pd_{x_2}\avg{u_1}$ in a range $0.2 \lesssim y/\delta \lesssim 0.7$. It is appealing to see that $\avg{C_{22}}$ is the component involved in the MDR dynamics, bearing in mind that $\avg{C_{11}} \gg \avg{C_{12}} \simeq \avg{C_{33}} > \avg{C_{22}}$. In the end, both Figs. \ref{fig:confab}b and \ref{fig:confscaling}a confirm the claims that $\avg{C_{12}}$ has reached its asymptotic limit within the Weissenberg numbers considered at this particular Reynolds number in this study, unlike $\avg{C_{22}}$ (see Figs. \ref{fig:confcd}a and \ref{fig:confscaling}b).

\section{Shear stress balance}
\label{sec:stressbalance}
The balance of total shear stress Eq. \eqref{eq:mombalancep} is considered in this section. The total shear stress in viscoelastic turbulent channel flow contains the shear stress $\beta\nu\frac{\dd}{\dd y}\avg{u}$ coming from the mean flow, the Reynolds stress $-\avg{u'v'}$ rising from turbulence and the mean polymer shear stress $\avg{\sigma_{12}}$ due to polymers in the flow, which is also referred to as the Reynolds stress deficit since $\nu\frac{\dd}{\dd y}\avg{u} - \avg{u'v'} \neq u_\tau^2 \lrbig{1 - y/\delta}$. In addition, the gradient of the Reynolds shear stress rises due to the non-linear term in the Navier-Stokes equations and it is related to the lift force (Magnus effect) experienced by a vortex line exposed to a velocity $\bm u$ \cite{tennekeslumley72}. The Reynolds shear stress and the rotational form $\bm u \times \bm \omega$ of the non-linear term in Eq. \eqref{eq:nondimNS2} are related through the following equation $\frac{\pd}{\pd y}\avg{u'v'} = \avg{v'\omega_z'} - \avg{w'\omega_y'}$. It is true that tangles of very intense and slender vortex filaments can exist down to very fine scales creating an energy drain on the mean flow. These tangles can be seen as one form of intermittency in turbulent flows \cite{frisch95}. However, it is unclear how the intermittency of the vorticify field $\bm \omega$ affects the averages $\avg{v'\omega_z'}$ and $\avg{w'\omega_y'}$ and thereby Reynolds shear stress via the relation $\frac{\pd}{\pd y}\avg{u'v'} = \avg{v'\omega_z'} - \avg{w'\omega_y'}$.

\begin{figure}[!ht]
 \includegraphics[width=8.5cm]{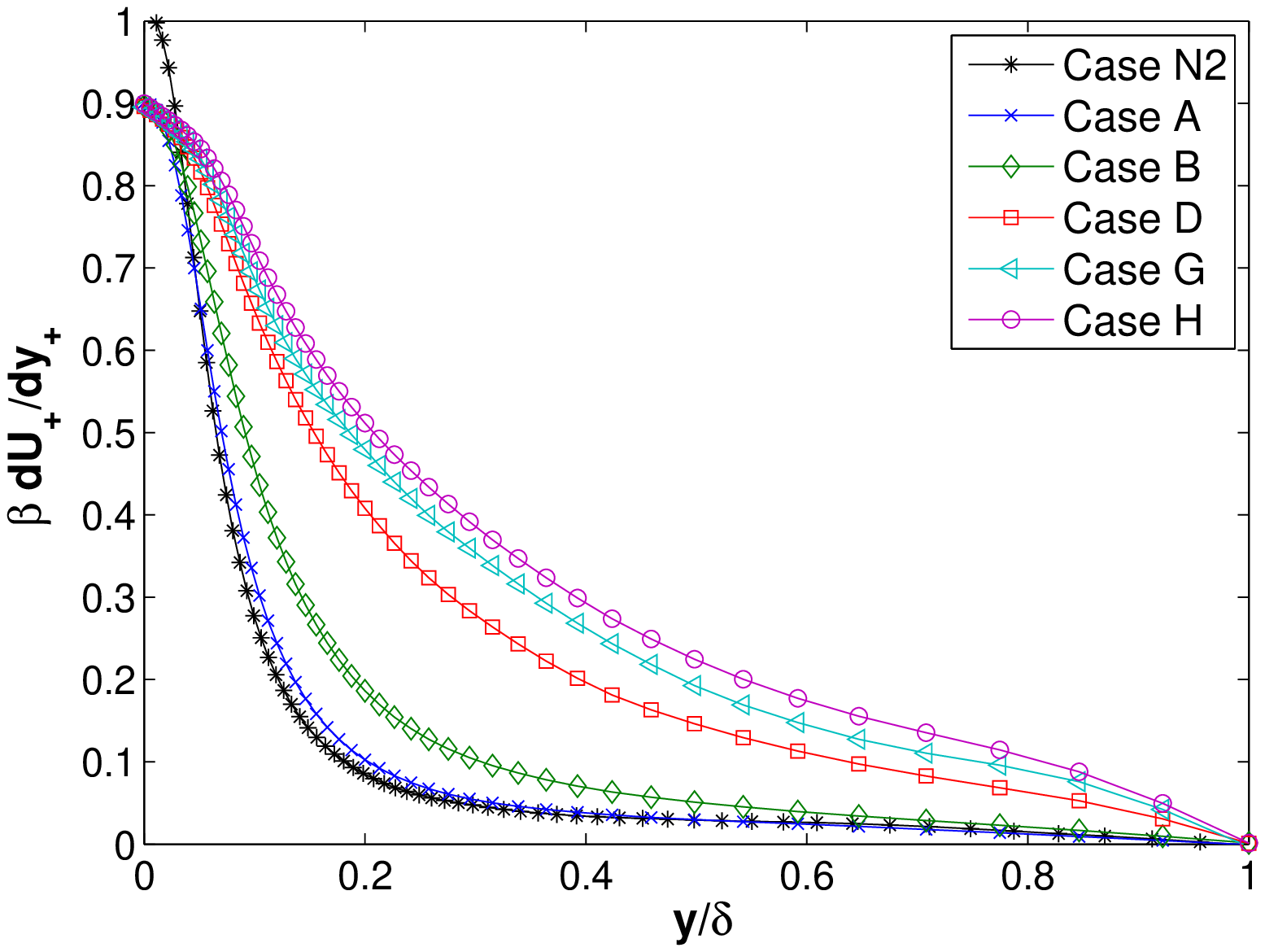}
 \includegraphics[width=8.5cm]{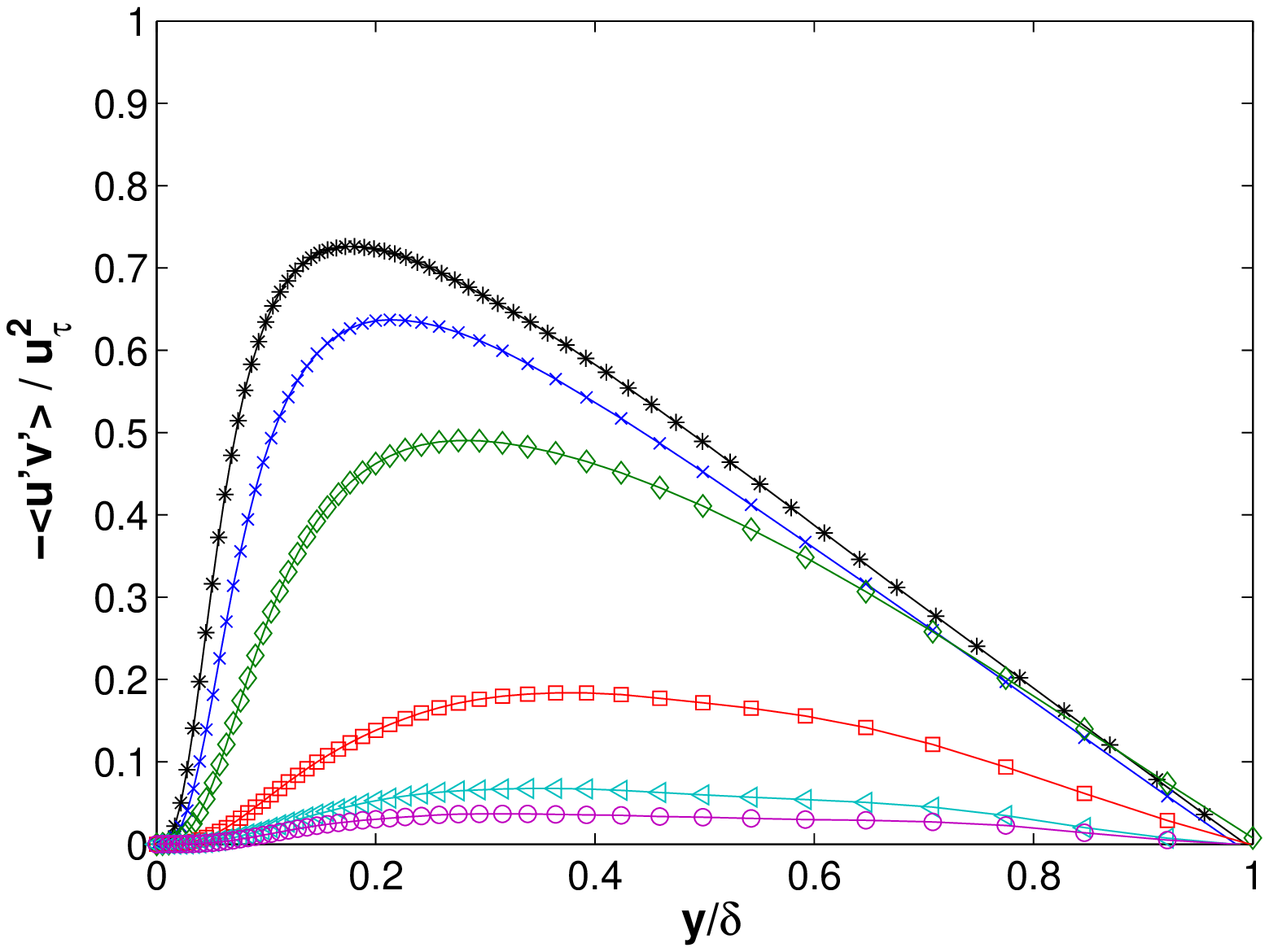}
 \includegraphics[width=8.5cm]{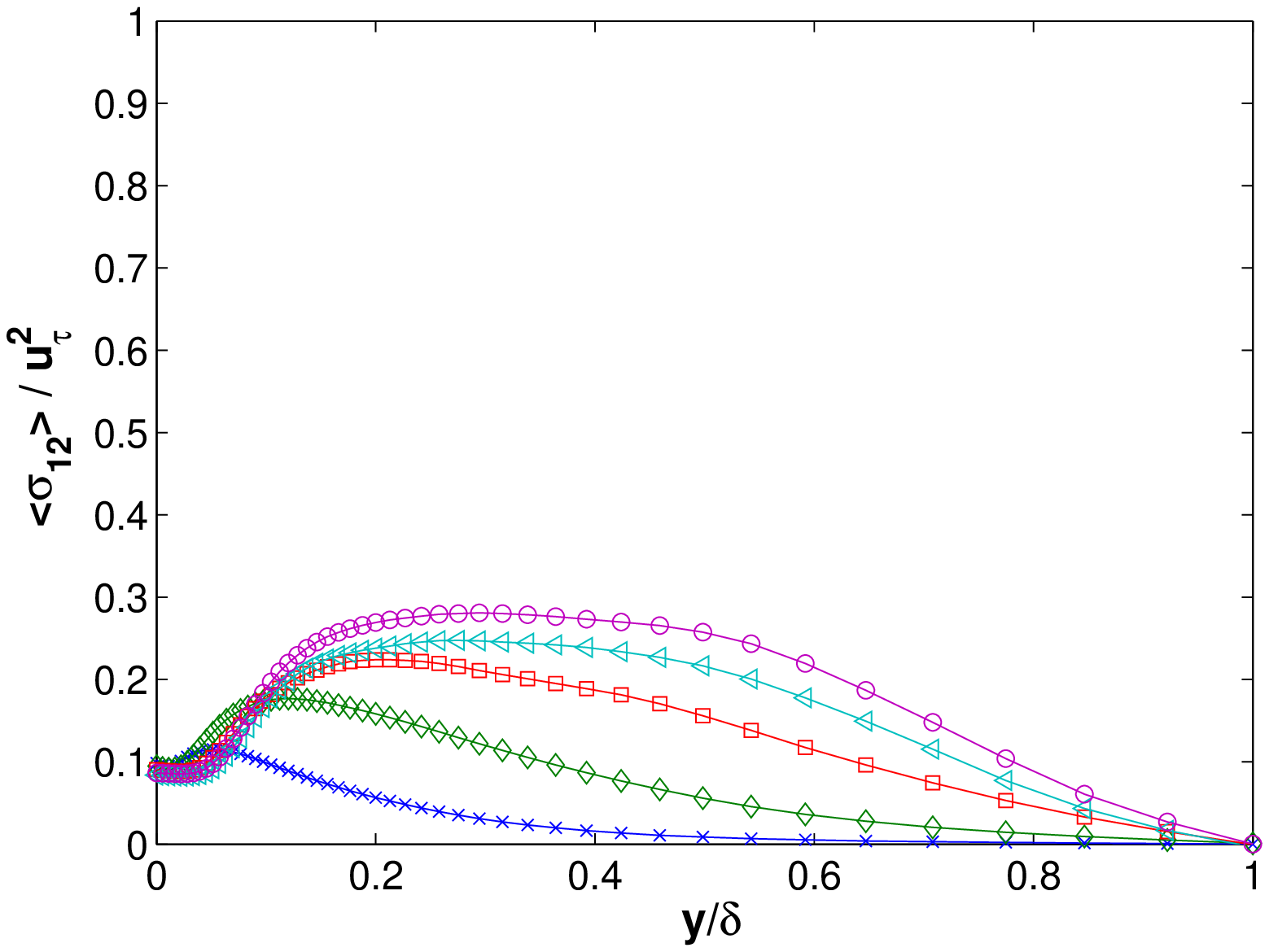}
 \caption{(Colour online) Profiles of (a) viscous shear stress, (b) Reynolds shear stress and (c) mean polymer shear stress versus $y/\delta$ for the LDR, HDR and MDR regimes. Note: case N2 ($\text{DR}=0\%$); case A ($\text{DR}=-14.2\%$); case B ($\text{DR}=-33.8\%$); case D ($\text{DR}=-57.3\%$);  case G ($\text{DR}=-62.1\%$); case H ($\text{DR}=-64.5\%$).}
 \label{fig:ymomentum}
\end{figure}

The viscous stress of the solvent, the Reynolds shear stress and the mean polymer shear stress normalised with viscous scales are presented in Fig. \ref{fig:ymomentum} at different levels of percentage DR for cases with the same $\text{Re}_c$ from Table \ref{tbl:dnspparameters}.
At the wall, the no-slip boundary condition enforces $-\avg{u'v'}\big|_{y = 0} = 0$. The wall shear stress is governed by $90\%$ viscous as well as $10\%$ polymer contribution for all viscoelastic cases as opposed to the Newtonian case N2. Viscosity is the dominant parameter in the near-wall region but becomes more influential in the outer regions as drag reduction enhances. This is clear from Fig. \ref{fig:ymomentum}a where $\beta \frac{\dd}{\dd y_+}U_+$ increases monotonically towards the centre of the channel as $\text{We}_c$ increases. Viscoelastic effects become also more significant towards the centre of the channel for higher $\text{We}_c$ cases. Reynolds shear stress, on the other hand, is constantly decorrelated at higher percentage DR with its peak shifting away from the wall. It is interesting to see that for the HDR case D $-\avg{u'v'}$ and $\avg{\sigma_{12}}$ are comparable and as MDR is approached the polymer shear stress plays an increasingly fundamental role in sustaining turbulence due to the vast attenuation of the Reynolds shear stress at these finite Reynolds number computations. This becomes apparent in the next section by analysing the turbulent kinetic energy budget.

Notice that Reynolds shear stress remains finite at MDR confirming the experimental measurements by Ptasinski \etal \cite{ptasinskietal01} against the complete depletion of $-\avg{u'v'}$ reported by Warholic \etal \cite{warholicetal99} and their subsequent claim that turbulence is sustained entirely by polymer stresses. What can be said theoretically on this controversy is the following. Consider first the limit of $\text{We}_S \to \infty$, where $A_2 \to 1$ for Eq. \eqref{eq:a2} even at finite Reynolds numbers, as Fig. \ref{fig:confscaling}b suggested. Then, the total shear stress balance Eq. \eqref{eq:mombalancep} can be rewritten using Eq. \eqref{eq:a2}
\begin{equation}
 \label{eq:stressbalance}
 \nu(\beta + (1-\beta)\avg{C_{22}})\frac{\dd\avg{u}}{\dd y} - \avg{u'v'} \simeq u_\tau^2 \lrbig{1 - \frac{y}{\delta}}
\end{equation}
where $\nu_{eff}(y) \equiv \nu(\beta + (1-\beta)\avg{C_{22}})$ is an effective viscosity similar to the one encountered in Lumley's phenomenology \cite{lumley73,procacciaetal08}. Now, when $\text{We}_S \gg 1$ assume that $\avg{C_{22}}$ becomes minimal based on theoretical claims by \cite{lvovetal05,procacciaetal08} and observational indications in this study. Then, for high enough Reynolds number along the universal MDR asymptotic line, i.e. taking first the infinite Weissenberg number limit and then the infinite Reynolds number limit, one might expect an intermediate region $\delta_\nu \ll y \ll \delta$ of approximately constant Reynolds shear stress, i.e. $-\avg{u'v'}/u_\tau^2 \to 1$, implied by Eq. \eqref{eq:stressbalance} when taking the limits of $y/\delta \to 0$ and $y/\delta_\nu \to \infty$ with the reasonable assumption that $\nu\beta\frac{\dd}{\dd y}\avg{u} \to 0$ as $y \gg \delta_\nu$. This statement suggests that the classical way of turbulence production does not vanish in the infinite Weissenberg and Reynolds number limit. 

Ultimately, the conjecture here is that $\avg{\sigma_{12}}$ becomes negligible in the stress balance Eq. \eqref{eq:mombalancep} when carefully taking the double Weissenberg and Reynolds number limit in the right order, so that we are on the universal MDR asymptotic line. This, however, does not indicate that drag reduction is depleted, it rather suggests that the MDR asymptote could be entirely determined by the energetics in these infinite limits. Nevertheless, polymers play a crucial role in the dynamics at MDR and this will be explored further in the next section.

\section{Polymer-turbulence dynamical interactions}
\label{sec:energybalance}
The balance equation for the turbulent kinetic energy of a viscoelastic fluid provides further insight into the dynamical interactions between polymers and turbulence. Assuming statistical stationarity and homogeneity in $x$ and $z$ directions for the mean turbulent kinetic energy balance and integrating over the $y$ direction, we obtain
\begin{equation}
 \label{eq:yintturbenergy}
 \int \mathcal P \,\dd y = \int \varepsilon_N \,\dd y + \int \varepsilon_P \,\dd y 
\end{equation}
with no contribution from the transport terms due to the no-slip boundary condition, using the divergence theorem. The turbulence production by Reynolds shear stress is denoted here by $\mathcal P \equiv -\avg{u'v'}\frac{\dd}{\dd y}\avg{u}$, the viscous dissipation rate $\varepsilon_N \equiv 2\nu\beta\avg{s_{ij}s_{ij}}$ and the viscoelastic dissipation rate $\varepsilon_P \equiv \avg{\sigma'_{ij}\pd_{x_j}u'_i}$, which arises due to fluctuating polymer stresses. Note that $\varepsilon_P$ has a dual nature, i.e. it can serve either as dissipation or production depending on the signs of the polymer stress fluctuations and that of the fluctuating velocity gradients.

\begin{figure}[!ht]
 \includegraphics[width=8.5cm]{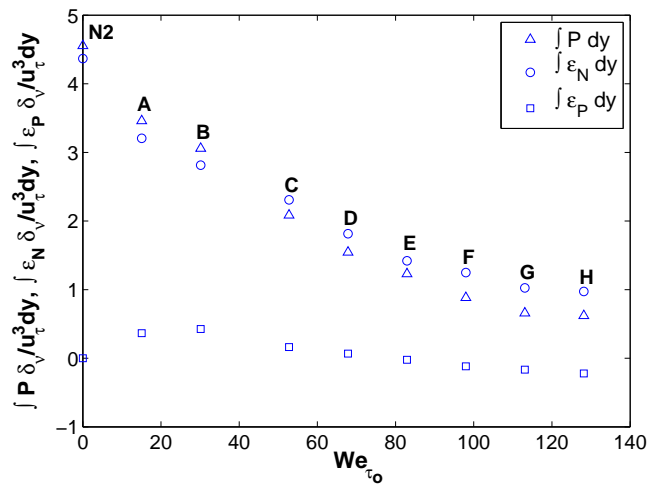}
 \caption{(Colour online) Terms of the $y$-integrated turbulent energy balance with respect to $\text{We}_{\tau_0}$.}
 \label{fig:intenergyterms} 
\end{figure}

Figure \ref{fig:intenergyterms} presents each term of Eq. \eqref{eq:yintturbenergy} normalised by $\delta_\nu / u_\tau^3$ with respect to $\text{We}_{\tau_0}$ for all cases from Table \ref{tbl:dnspparameters} at $\text{Re}_c = 4250$.
An asymptotic behaviour to a marginal flow state can be observed by increasing the polymer relaxation time scale with a vast attenuation occurring in the total production and viscous dissipation, while viscoelastic dissipation grows mildly in the LDR regime and constantly decays within HDR and MDR. Overall, $\int \varepsilon_P \,\dd y$ becomes pivotal in the dynamics of the flow relative to $\int \mathcal P \,\dd y$ and $\int \varepsilon_N \,\dd y$ for HDR and MDR flows. Most importantly $\int \varepsilon_P \,\dd y < 0$ for high $\text{We}_{\tau_0}$ values according to Fig. \ref{fig:intenergyterms}, in agreement with experimental measurements \cite{ptasinskietal01}, implying that polymers somehow can sustain turbulence by producing turbulent kinetic energy. Notice, that in this plot both dissipations are presented as positive quantities and this was done on purpose to emphasise the interplay between production and viscous dissipation from LDR to HDR. It is noteworthy that $\int \mathcal P \,\dd y > \int \varepsilon_N \,\dd y$ for LDR cases A and B but $\int \mathcal P \,\dd y < \int \varepsilon_N \,\dd y$ for HDR cases and gets even smaller as drag reduction approaches its maximum limit. This observation hints that polymer dynamics get somehow involved in the production of turbulent kinetic energy so that turbulence does not die out at HDR and MDR.

\begin{figure}[!ht]
   \includegraphics[width=8.5cm]{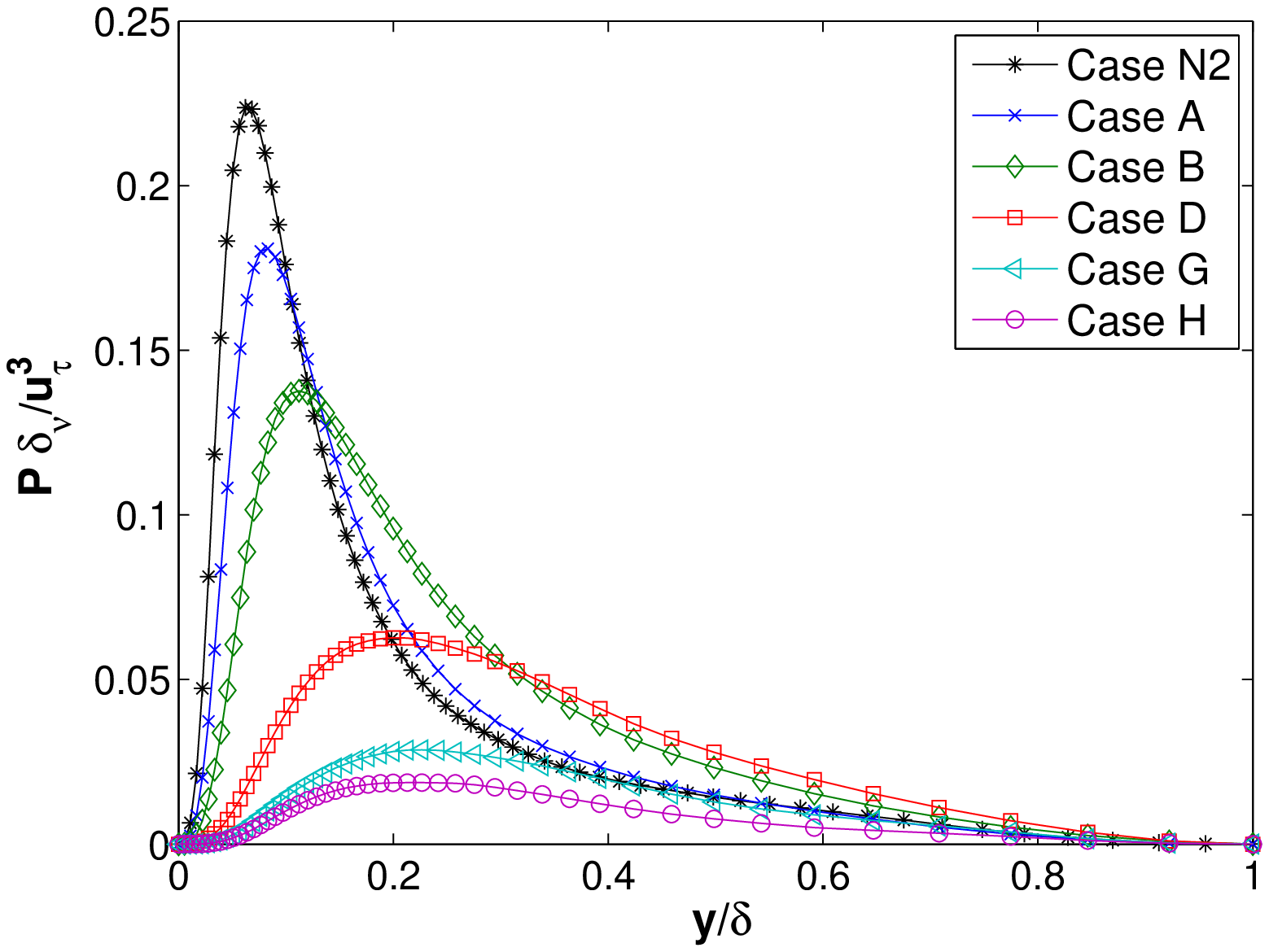}
   \includegraphics[width=8.5cm]{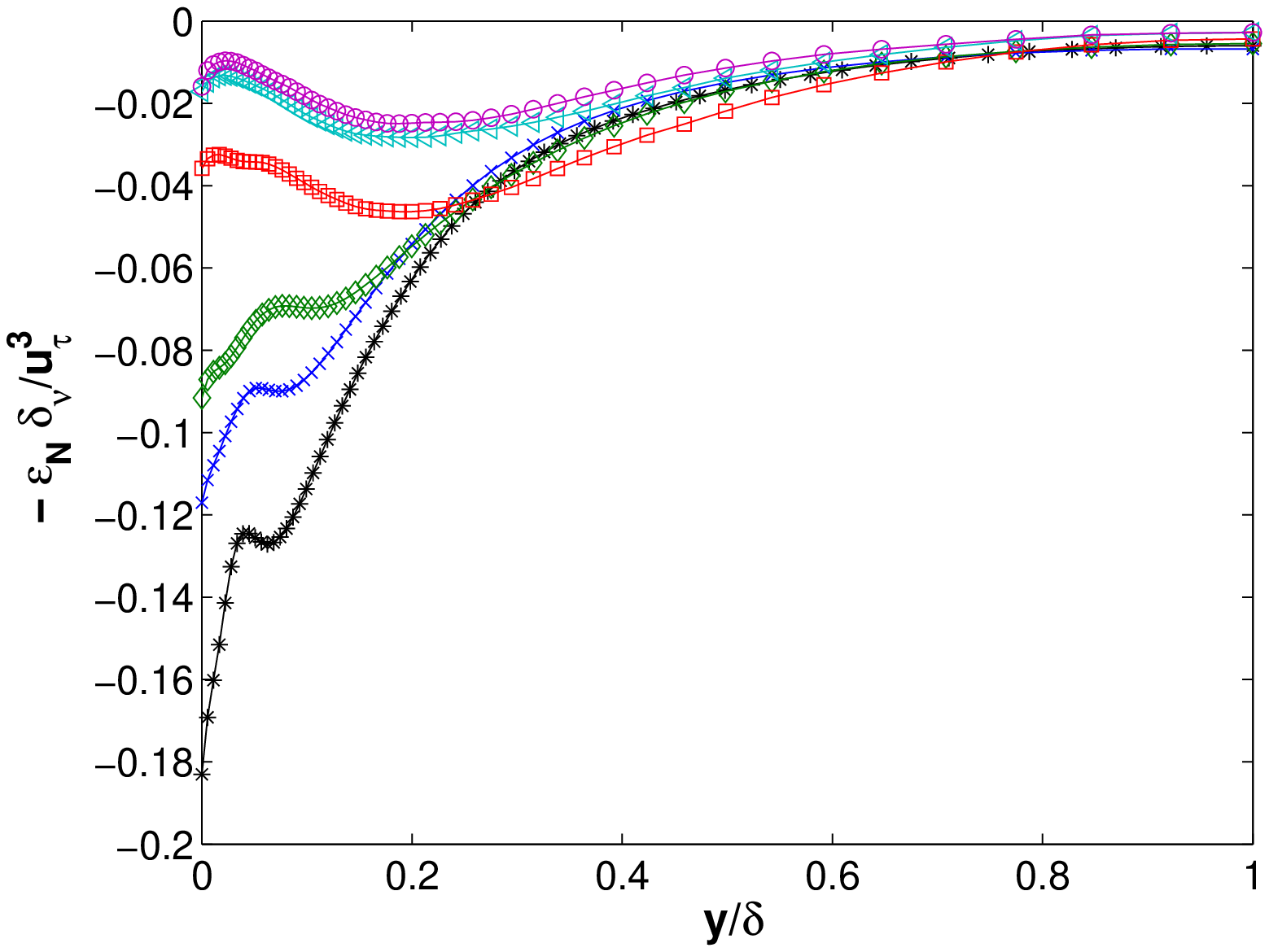}
   \includegraphics[width=8.5cm]{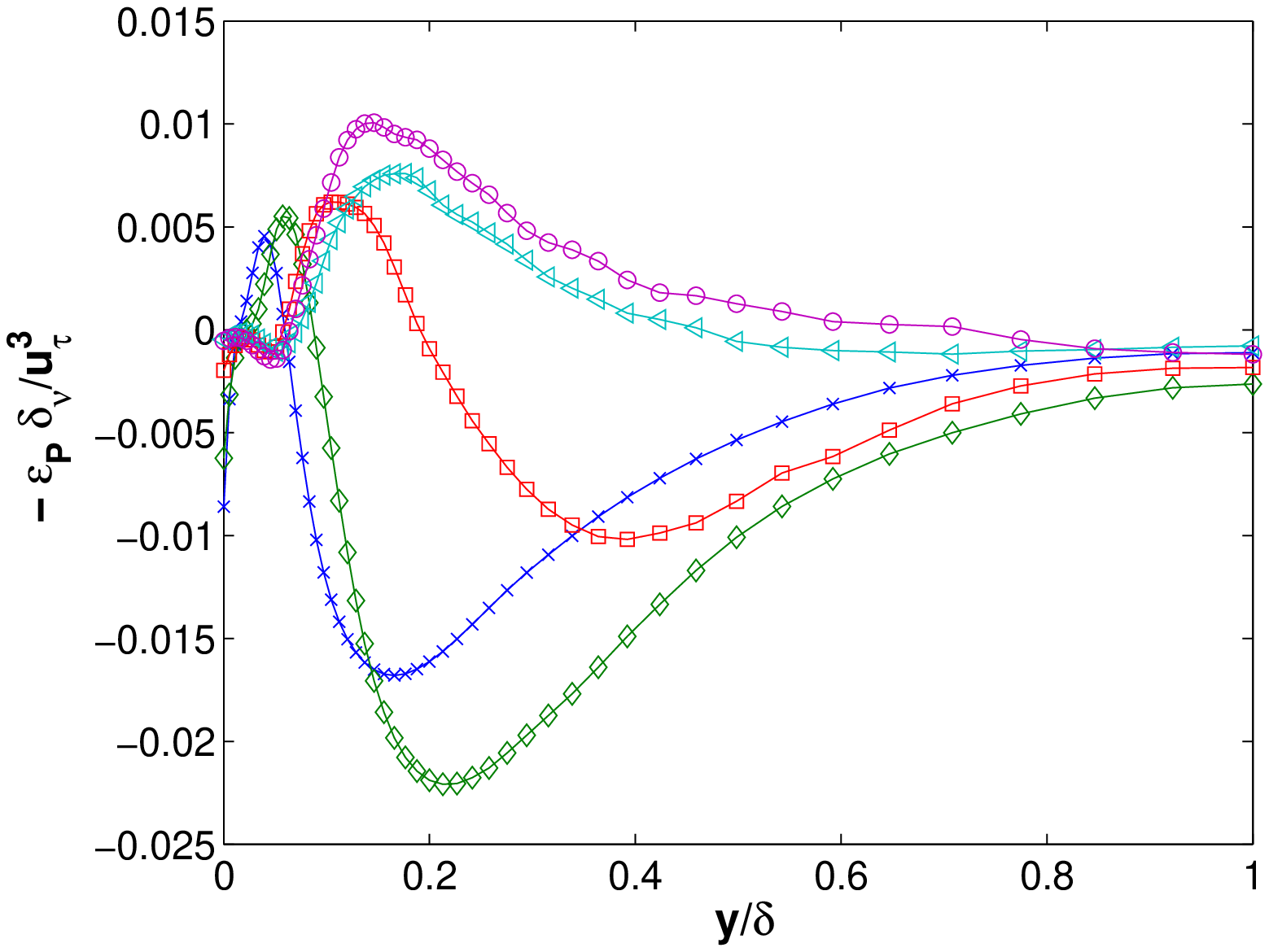}
  \caption{(Colour online) Profiles of (a) turbulence production $\mathcal P \delta_\nu/u_\tau^3$, (b) viscous dissipation $-\varepsilon_N \delta_\nu/u_\tau^3$ and (c) viscoelastic dissipation $-\varepsilon_P \delta_\nu/u_\tau^3$ for the LDR, HDR and MDR. Note: case N2 ($\text{DR}=0\%$); case A ($\text{DR}=-14.2\%$); case B ($\text{DR}=-33.8\%$); case D ($\text{DR}=-57.3\%$);  case G ($\text{DR}=-62.1\%$); case H ($\text{DR}=-64.5\%$).}
  \label{fig:energyterms}
\end{figure}

Lets now look in more detail at the profiles of $\mathcal P$, $\varepsilon_N$ and $\varepsilon_P$ scaled by $\delta_\nu/u_\tau^3$ with respect to normalised distance from the wall $y/\delta$ for representative cases at various levels of drag reduction from Table \ref{tbl:dnspparameters} (see Fig. \ref{fig:energyterms}). Dissipation represents drain of energy, hence, $\varepsilon_N$ and $\varepsilon_P$ have been plotted here as negative quantities.

The production of turbulent energy by Reynolds stresses, which is continuously reduced over the extend of drag reduction as a function of $\text{We}_c$, serves to exchange kinetic energy between the mean flow and the turbulence. The local peak of $\mathcal P$ is reached within the buffer layer and in fact for Newtonian flows we can easily show that the maximum production occurs where $-\avg{u'v'} = \nu \frac{\dd}{\dd y}\avg{u}$ and $\mathcal P_{max}\delta_\nu / u_\tau^3 < \frac{1}{4}$ \cite{pope00}. The peak turbulence production within the LDR regime also occurs at the intersection point of viscous and Reynolds shear stress (compare Figs. \ref{fig:ymomentum}a and \ref{fig:ymomentum}b with Fig. \ref{fig:energyterms}a), which shifts away from the wall as $\text{We}_c$ increases, indicating the thickening of the elastic layer. However, for HDR and MDR cases $\mathcal P_{max}\delta_\nu / u_\tau^3$ is within $0.1 < y/\delta \lesssim 0.3$, where the maximum Reynolds stress roughly appears, without following the $-\avg{u'v'} = \beta\nu \frac{\dd}{\dd y}\avg{u}$ intersection point, which does not even exist for cases G and H (see Figs. \ref{fig:ymomentum}a and \ref{fig:ymomentum}b).

Viscous dissipation exhibits monotonic attenuation as drag reduces for higher values of $\text{We}_c$ with the maximum dissipation arising at the wall for the Newtonian case N2 and the LDR cases A and B (see Fig. \ref{fig:energyterms}b). Although the kinetic energy is zero at the wall since $\bm u'|_{y=0} = 0$ imposed by the no-slip boundary conditions, the fluctuating strain rate and consequently $\varepsilon_N$ is non-zero. At high percentage DR, we surprisingly observe that the highest fluctuating strain rates are encountered away from the wall providing a completely different picture of the near-wall dissipation dynamics. The local kink in the buffer/elastic layer, which arises due to intense activity in this region, exists at corresponding $y/\delta$ with $\mathcal P_{max}\delta_\nu / u_\tau^3$ for all cases considered in Fig. \ref{fig:energyterms}b and becomes a global minimum for the HDR and MDR cases, dominating the profiles of viscous dissipation.

\begin{figure}[!ht]
 \includegraphics[width=8.5cm]{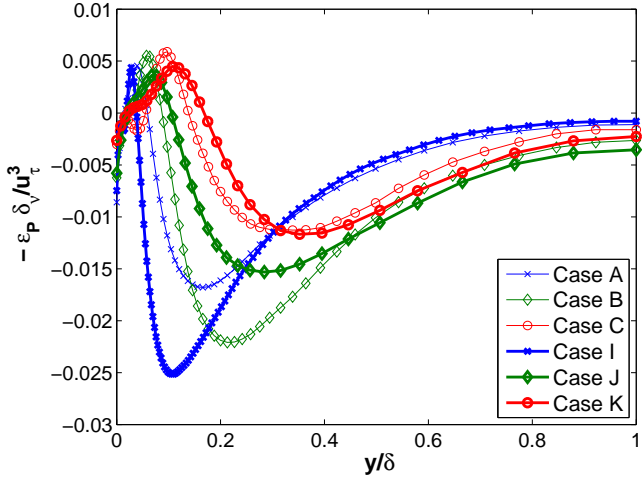}
 \caption{(Colour online) Effect of Reynolds number on viscoelastic dissipation as function of $y/\delta$. Identical symbols correspond to cases with the same $\text{We}_c$ values. Compare case A ($\text{We}_c = 2$, $\text{Re}_c = 4250$) with case I ($\text{We}_c = 2$, $\text{Re}_c = 10400$); case B ($\text{We}_c = 4$, $\text{Re}_c = 4250$) with case J ($\text{We}_c = 4$, $\text{Re}_c = 2750$); case C ($\text{We}_c = 7$, $\text{Re}_c = 4250$) with case K ($\text{We}_c = 7$, $\text{Re}_c = 2750$).}
 \label{fig:disspReeffect} 
\end{figure}

The profiles of viscoelastic dissipation obey a characteristic transitional trend similar to what has been already observed for $u'_+$ (see Fig. \ref{fig:velrms}a) and $\avg{C_{22}}$ (see Fig. \ref{fig:confcd}a) from LDR to HDR regime, as $\text{We}_c$ increases. In detail, the curves of LDR cases A and B shift downwards increasing viscoelastic dissipation but those of the HDR/MDR cases move upwards enhancing the positive nature of $-\varepsilon_P \delta_\nu / u_\tau^3$. The dual nature of $\varepsilon_P$ is clearly depicted in Fig. \ref{fig:energyterms}c with polymers dissipating and producing turbulent kinetic energy in different regions, which depend on the polymer relaxation time scale at a given Reynolds number. A Reynolds number dependence of these regions is expected owing to the effect of different flow time scales on dumbbells with a particular relaxation time scale. Figure \ref{fig:disspReeffect} compares cases of identical Weissenberg numbers and different Reynolds numbers (see Table \ref{tbl:dnspparameters}), illustrating a weaker $\text{Re}_c$ dependence on viscoelastic dissipation in comparison to the stronger $\text{We}_c$ dependence in Fig. \ref{fig:energyterms}c, particularly at HDR and MDR.
The part of the total dissipation that occurs in the three regions defined by the profile of viscoelastic dissipation in Fig. \ref{fig:energyterms}c can be estimated based on the profiles in Figs. \ref{fig:energyterms}b and \ref{fig:energyterms}c. Approximately 15\% to 25\% of the total dissipation takes place in the first region, 25\% to 60\% in the second region and 20\% to 70\% in the third region. In other words, the majority of the total dissipation occurs away from the wall.

\begin{figure}[!ht]
  \includegraphics[width=8.5cm]{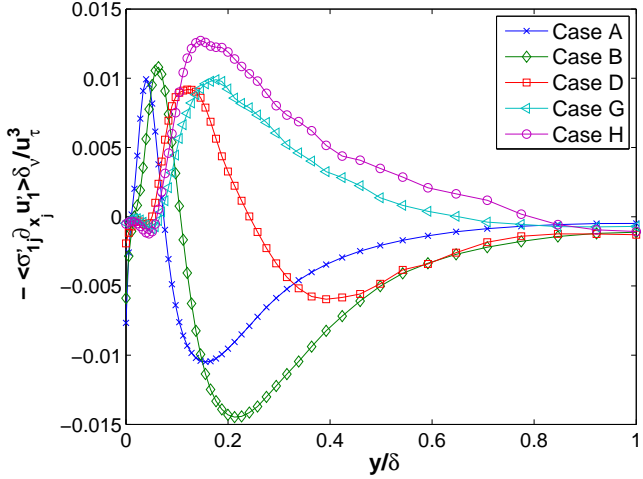}
  \caption{(Colour online) $-\avg{\sigma'_{1j}\pd_{x_j}u'_1}\delta_\nu/u_\tau^3$ as function of $y/\delta$ for the LDR, HDR and MDR regimes. Note: case A ($\text{DR}=-14.2\%$); case B ($\text{DR}=-33.8\%$); case D ($\text{DR}=-57.3\%$);  case G ($\text{DR}=-62.1\%$); case H ($\text{DR}=-64.5\%$).}
  \label{fig:epspa}
\end{figure}

Now, considering each component of the correlation matrix $\varepsilon_P \equiv \avg{\sigma'_{ij}\pd_{x_j}u'_i}$, where summation applies over the indices $i$ and $j$, we can observe that components with $i = 2, 3$ can be ignored, with most of the contribution ascribed to $i = 1$ components according to Fig. \ref{fig:epspa} which is very similar to Fig. \ref{fig:energyterms}c.
\begin{figure}[!ht]
 \includegraphics[width=8.5cm]{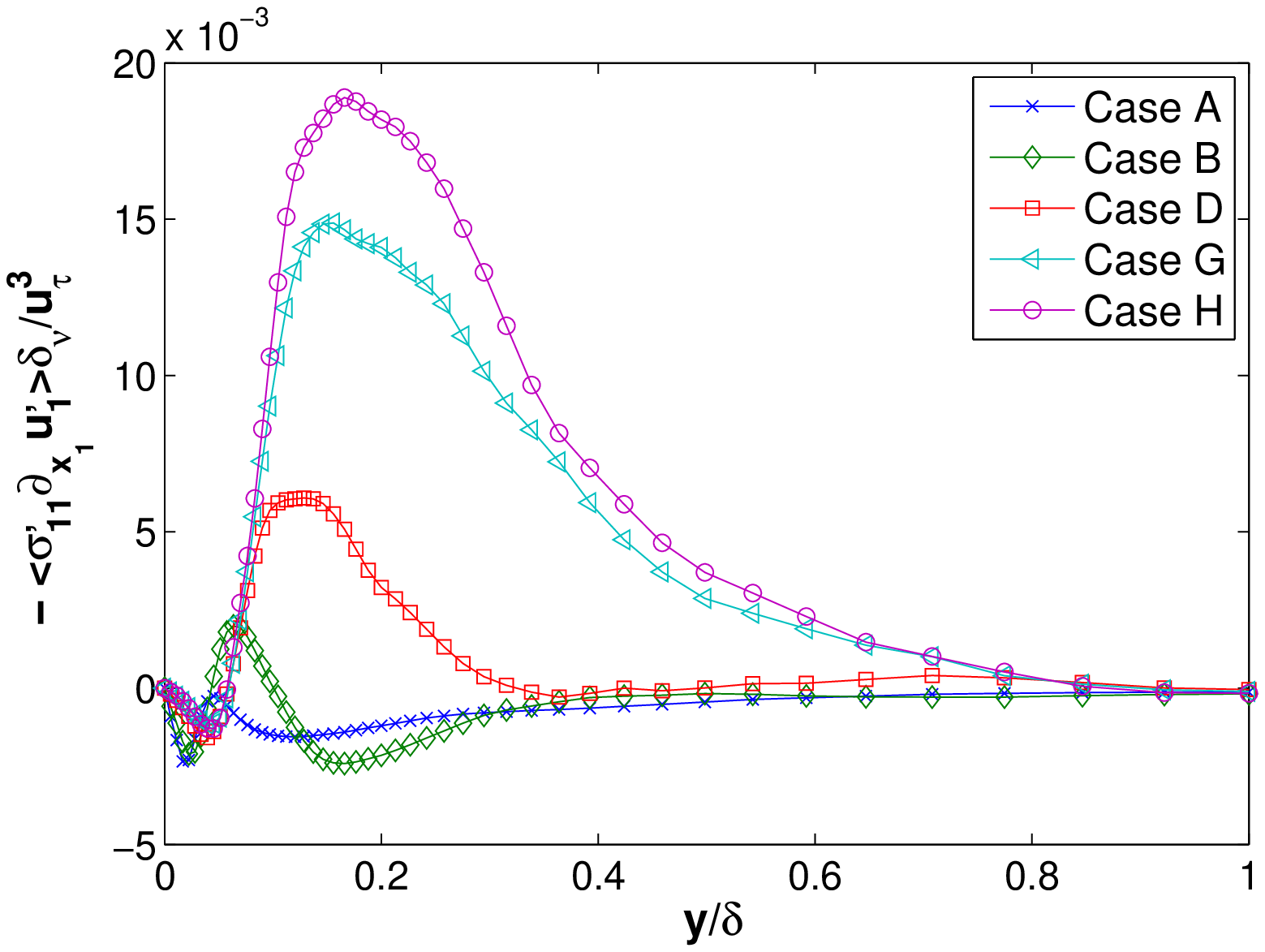}
 \includegraphics[width=8.5cm]{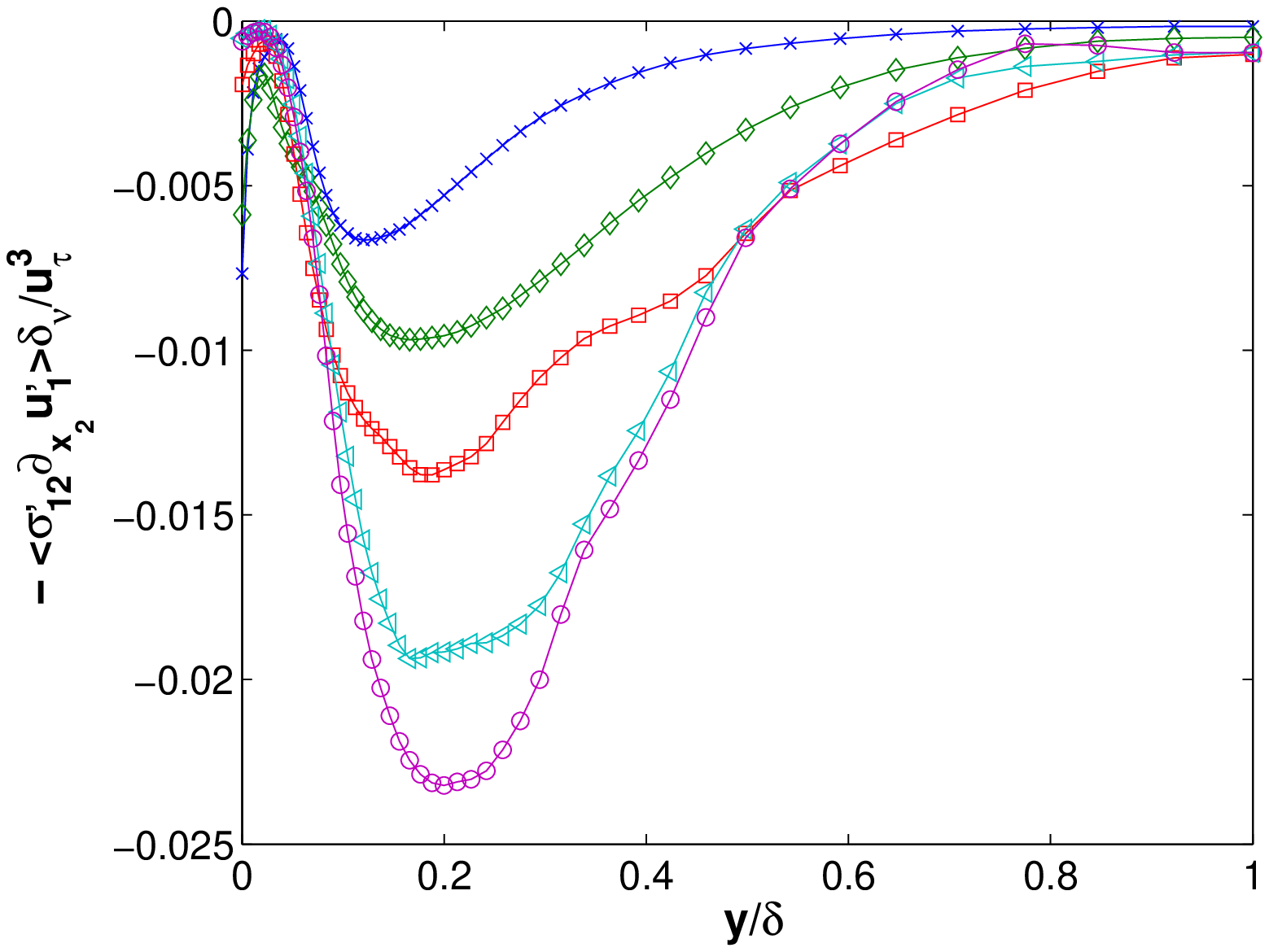}
 \includegraphics[width=8.5cm]{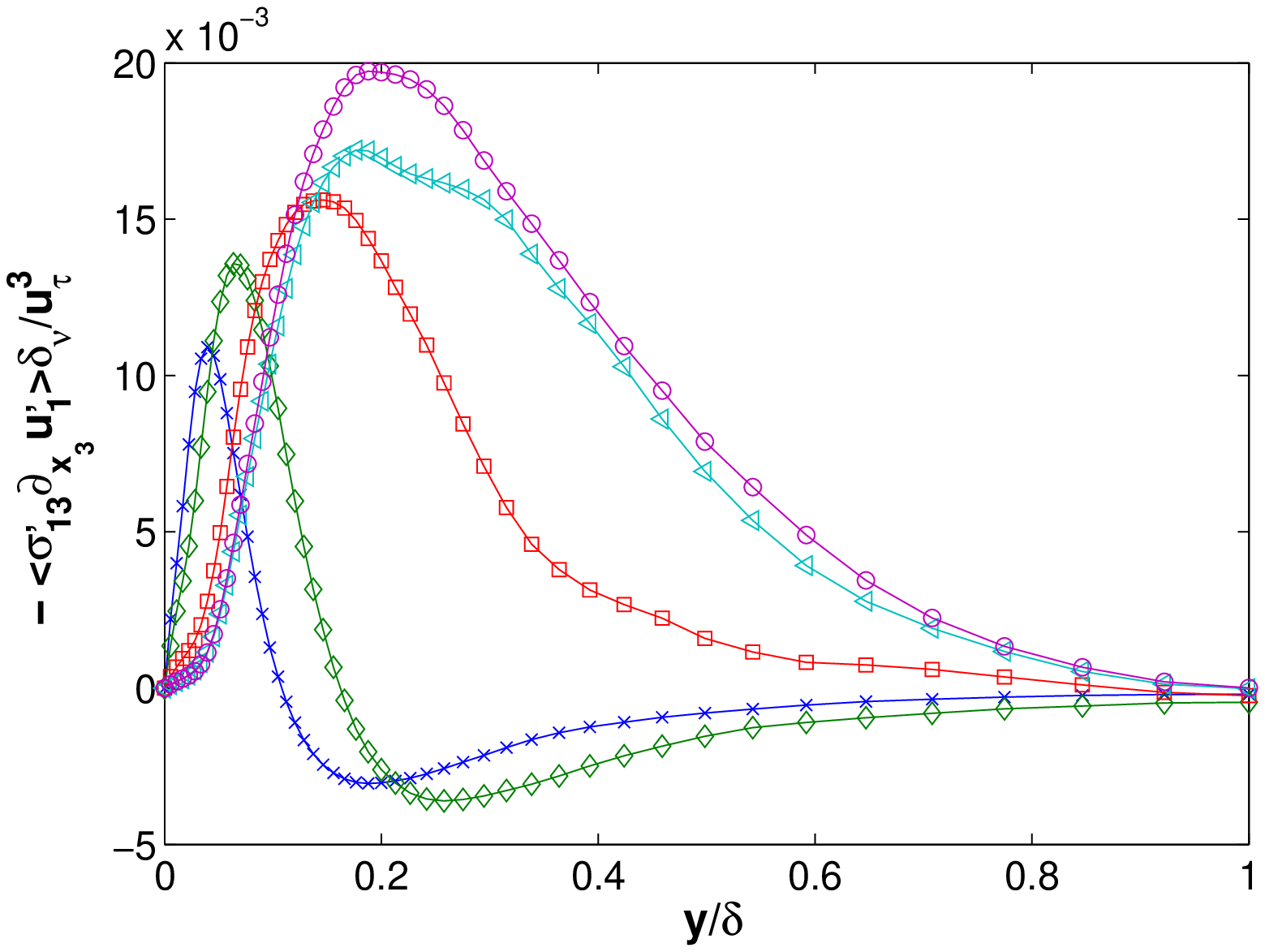}
 \caption{(Colour online) Profiles of viscoelastic dissipation components for the LDR, HDR and MDR regimes. (a) $-\avg{\sigma'_{11}\pd_{x}u'}\delta_\nu/u_\tau^3$, (b) $-\avg{\sigma'_{12}\pd_{y}u'}\delta_\nu/u_\tau^3$ and (c) $-\avg{\sigma'_{13}\pd_{z}u'}\delta_\nu/u_\tau^3$ as function of $y/\delta$. Note: case N2 ($\text{DR}=0\%$); case A ($\text{DR}=-14.2\%$); case B ($\text{DR}=-33.8\%$); case D ($\text{DR}=-57.3\%$);  case G ($\text{DR}=-62.1\%$); case H ($\text{DR}=-64.5\%$).}
 \label{fig:epspbcd}
\end{figure}
The qualitative features of $\varepsilon_P$ are clearly captured by $\avg{\sigma'_{1j}\pd_{x_j}u'_1}$, simplifying the underpining dynamics of viscoelastic dissipation. However, to be precise, $\varepsilon_P$ is neither exactly approximate nor proportional to $\avg{\sigma'_{1j}\pd_{x_j}u'_1}$. Note that the positive nature of $\varepsilon_P$ is caused by the correlations $-\avg{\sigma'_{11}\pd_{x_1}u'_1}$ and $-\avg{\sigma'_{13}\pd_{x_3}u'_1}$ (see Figs. \ref{fig:epspbcd}a and \ref{fig:epspbcd}c). The rest of the components are negative for all cases considered here and decrease monotonically as $\text{We}_c$ increases like $-\avg{\sigma'_{12}\pd_{x_2}u'_1}$ in Fig. \ref{fig:epspbcd}b. The only exception though is $-\avg{\sigma'_{32}\pd_{x_2}u'_3}$, which also exhibits a dual trend, negligible however in comparison to the components presented in Fig. \ref{fig:epspbcd}. Finally, the correlations in Figs. \ref{fig:epspbcd}a and \ref{fig:epspbcd}c are also responsible for the transitional behaviour of viscoelastic dissipation profiles from LDR to HDR discussed earlier.

The current picture of the dual nature of $\varepsilon_P$ was first predicted by Min \etal \cite{minetal03a} at low Weissenberg numbers, adding an artificial diffusion term to numerically solve the FENE-P model. However, the present DNS are the first to capture so clearly these regions throughout the drag reduction regimes, predicting the appropriate dynamics at corresponding $\text{We}_c$ values. Once more, this is attributed to our numerical approach applied here for the FENE-P model that is able to capture stronger polymer-turbulence interactions than algorithms based on artificial diffusion. There are even results using the artificial diffusion methodologies that erroneously predict polymers never feeding energy back to the flow \cite{ptasinskietal03,procacciaetal08}. Hence, in view of the current distinctly transparent observations a conceptual model for the mechanism of drag reduction is deduced in the next section.

\section{Drag reduction mechanism}
\label{sec:DRmechanism}
The recent review on polymer drag reduction by White and Munghal \cite{whitemunghal08} reports that the numerical evidence is somewhat conflicting regarding the flow regions where polymers extend and contract. In this study, these regions can be identified by applying the Reynolds decompositions $u_i = \avg{u_i} + u_i'$ and $\sigma_{ij} = \avg{\sigma_{ij}} + \sigma_{ij}'$ to Eq. \eqref{eq:Epeqn}, following the spirit of \cite{minetal03a,ptasinskietal03}. Then, we can notice that $\avg{\sigma'_{ij}\pd_{x_j}u'_i}$ appears as a production term due to turbulence for the mean polymer elastic energy. Now, from the definition of polymer elastic energy Eq. \eqref{eq:elasticenergy}, it is evident that $\avg{E_p} \propto \ln\avg{f(C_{kk})}$. So, the FENE-P dumbbells are stretched when $-\varepsilon_P \delta_\nu / u_\tau^3 < 0$ in Fig. \ref{fig:energyterms}c and then elastic energy is stored on polymers, absorbing turbulent kinetic energy from the flow. Hence, a mechanism of drag reduction can be proposed based on the polymers stretching or in other words the behaviour of viscoelastic dissipation as a function of the distance $y$ from the wall.

According to Figs. \ref{fig:energyterms}c and \ref{fig:disspReeffect} there are three main regions in the profiles of viscoelastic dissipation
\begin{equation}
 \label{eq:disspdivision}
  -\varepsilon_P \delta_\nu / u_\tau^3
 \begin{cases}
  < 0 \,, & 0 \leq y/\delta < \delta_1(\text{We}_c,\text{Re}_c) \\
  > 0 \,, & \delta_1(\text{We}_c,\text{Re}_c) \leq y/\delta \leq \delta_2(\text{We}_c,\text{Re}_c) \\
  < 0 \,, & \delta_2(\text{We}_c,\text{Re}_c) < y/\delta \leq 1.
 \end{cases}
\end{equation}
The first region is at the proximity of the wall, where polymers unravel because of the high mean shear, consistent with other studies \cite{massahhanratty97,minetal03a,terraponetal04,dubiefetal04}, storing elastic potential energy. The range of this region has a weak dependence on Weissenberg and Reynolds number with its upper bound being within the viscous sublayer $\delta_1(\text{We}_c,\text{Re}_c) \lesssim 0.05$ for all $\text{We}_c$ and $\text{Re}_c$ cases considered. 

The second region is the most interesting since polymers release energy back to the flow, contracting towards their equilibrium length, as they are convected away from the wall by the near-wall vortical motions. The manifestation of turbulence production by polymers can be interpreted in terms of the correlation of the polymers with the local fluctuating strain rates and their persistence in this region. In particular, $-\avg{\sigma'_{ij}\pd_{x_j}u'_i}$ reveals that $-\avg{\sigma'_{11}\pd_{x} u'}$ as well as $-\avg{\sigma'_{13}\pd_{z} u'}$ are responsible for the contraction of the dumbbells and consequently for the release of the stored elastic energy, since they are positively correlated in this region away from the wall (see Figs. \ref{fig:epspbcd}a and \ref{fig:epspbcd}c). This region exists in an intermediate $y/\delta$ range, whose upper bound $\delta_2(\text{We}_c,\text{Re}_c)$ is strongly dependent on $\text{We}_c$ and less on $\text{Re}_c$ values. As drag reduction amplifies for larger polymer relaxation time scales this positive region expands to a wider $y/\delta$ range, which dominates the nature of $-\varepsilon_P \delta_\nu / u_\tau^3$ at MDR (see Fig. \ref{fig:energyterms}c). 

Finally, polymers transported away from the wall get also negatively correlated with the persistent fluctuating strain rates (see Fig. \ref{fig:epspbcd}b) and are extended in the region $\delta_2(\text{We}_c,\text{Re}_c) < y/\delta \leq 1$, which is a sink for turbulent kinetic energy, prevailing the LDR flows. However, this region is diminished for HDR and MDR flows (see Fig. \ref{fig:energyterms}c) due to the interplay between the productive and dissipative inherent features of $\varepsilon_P$, which mainly depend on the polymer relaxation time scale and the existence of intense velocity fluctuations that are able to stretch the polymer molecules.

The phenomenology of the proposed mechanism shares many similarities with various conceptual models of earlier works \cite{whitemunghal08}. In this study, the basic idea is that the transport of the elastic potential energy, stored by polymers near the wall, is mainly associated with the polymer relaxation time scale. The latter determines the distribution of energy away from the wall and as a consequence the near-wall turbulence dynamics weaken. Up to this point, the mechanism 
agrees with the interpretation of Min \etal \cite{minetal03a}, which is essentially confirmed by the present illustrative computations. However, the novelty here is that this mechanism is valid for higher $\text{We}_c$ values and levels of percentage DR in contrast to Min \etal \cite{minetal03b}, who claim that it is not valid for HDR/MDR flows basing their arguments on their debatable numerical results (see also section \ref{sec:velstats}). 

In addition, the refinement of the proposed conceptual mechanism resides on the reduction of $\varepsilon_P$ to $\avg{\sigma'_{1j}\pd_{x_j}u'_1}$ and even more on the correlations $\avg{\sigma'_{11}\pd_{x}u'}$ and $\avg{\sigma'_{13}\pd_{z}u'}$, which are responsible for the turbulence production by polymer coils. The existence of a third dissipative region away from the wall is also emphasised in this mechanism, where polymers, after their contraction, are now stretched by the intense fluctuating velocity field. This outer region dominates the viscoelastic dissipative dynamics of the LDR regime and diminishes asymptotically as $\text{We}_c$ increases but it never disappears. Ultimately, this picture along with the anisotropy introduced into the components of turbulent kinetic energy, i.e. $E = \frac{1}{2}(u'^2 + v'^2 + w'^2)$, comprise the drag reduction mechanism deduced in this study.
\section{Conclusion}
\label{sec:conclusion}
This paper is devoted to the polymer dynamics in viscoelastic turbulent channel flow and their effects on the flow, reproducing turbulent drag reduction by DNS using a state-of-the-art numerical scheme in wall-bounded flows to solve the FENE-P model. The potential of this methodology to capture the strong polymer-turbulence dynamical interactions allowed $\beta$ values to remain high, more representative of dilute polymer solutions used in experiments. Even then, higher percentage DR values are obtained for given $\text{We}_c$ than previous numerical studies.

The effects of $L_p$ and $\text{Re}_c$ on the results support the claims for non-universality of the dynamics for intermediate levels of DR between the von K\'arm\'an and the MDR law. The universal MDR asymptote, on the other hand, is reached in this study under the combination of high polymer extensibility $L_p$ with high enough elasticity given by large values of $\text{We}_c$ at a given moderate $\text{Re}_c$.

The experimentally observed distinct differences in the statistical trends of the turbulent velocity field, particularly for $u'_+$ (see Fig. \ref{fig:velrms}a), are clearly identified with the current numerical approach in comparison with other simulations, most of which do not even approach such a characteristic trend. Overall, the peaks of the statistical profiles of velocity and vorticity fluctuations shift away from the wall as DR increases, in agreement with other experimental and numerical studies, indicating the thickening of the buffer layer. At the same time, $\nu\beta\frac{\dd}{\dd y}\avg{u}$ increases towards the centre of the channel for higher $\text{We}_c$, denoting the importance of viscosity away from the wall at these moderate Reynolds number DNS.

Lumley's phenomenology \citet{lumley69} on the manifestation of drag reduction is based on the conjecture of coil-stretch transition, i.e. exponential full uncoiling of polymer molecules, for the build-up of intrinsic viscosity. However, our numerical results illustrate that the onset of drag reduction and even the MDR asymptotic state can be reached while $\avg{C_{kk}} \ll L_p^2$ with $L_p$ large enough, in agreement with the initial claim by Tabor and de Gennes \citet{tabordegennes86} that even high space-time strain rate fluctuations near the wall can only partially stretch polymer coils. We also showed that the percentage polymer extension is less but the actual extension is more for larger $L_p$, amplifying DR. Thus, large polymer coils that do not reach their critical full extensibility should be of interest to experimental investigations on scission degradation of polymer chains and drag reduction effectiveness. Such macromolecules would be less vulnerable to rupture avoiding the loss of the drag reduction effect.
Besides, they should be able to stretch substantially to make a stronger impact on turbulent activity and consequently enhance percentage drag reduction.

The analysis of the conformation tensor field provides great insight into the polymer dynamics and their influence on the flow. The dominant anisotropic behaviour of the mean conformation tensor, i.e. $\avg{C_{11}} \gg \avg{C_{12}} \simeq \avg{C_{33}} > \avg{C_{22}}$, due to the mean shear in viscoelastic turbulent channel flow, influences the anisotropy of the fluctuating flow field. The anisotropy in the HDR and MDR regimes is depicted at the small scales of our DNS outside the buffer layer and towards the centre of the channel by $\omega'_{z_+} > \omega'_{y_+} > \omega'_{x_+}$. 

Different asymptotic rates of convergence are observed for the conformation tensor components towards the limit of infinite Weissenberg number demonstrating the complex polymer dynamics even in this simplified dumbbell model. In the limit $\text{We}_S \to \infty$ polymers are considered stiff, i.e. $C_{ij} \to \avg{C_{ij}}$, mostly in the main directions of elongation and the correlations of the fluctuating conformation tensor and velocity fields in the other directions are assumed to remain minimal at this limit. Therefore, $\avg{\sigma_{11}} = A_1 \frac{1-\beta}{\text{Re}_c} 2\avg{C_{12}}\frac{\dd}{\dd y}\avg{u}$ and $\avg{\sigma_{12}} = A_2 \frac{1-\beta}{\text{Re}_c} \avg{C_{22}}\frac{\dd}{\dd y}\avg{u}$, with $A_1 \to 1$ and $A_2 \to 1$ in a region somewhere between the wall and the centre of the channel in that limit. Our numerical results show that $A_1 \to 1$ in such a region but not $A_2$. $A_2$ on the other hand is about contant in the range $0.2 \lesssim y/\delta \lesssim 0.7$ and shows a tendency towards 1 as $\text{We}_S$ increases.

The following theoretical view was stated in this paper with regards to the controversy over the existence or not of Reynolds shear stress at the MDR limit, which is of fundamental importance to the dynamics of turbulence production at this limit. It is conjectured that at the MDR limit $\avg{\sigma_{12}}$ is negligible. This was based on the idea mentioned above about the stiffness of polymers at $\text{We}_S \to \infty$ plus the assumption that $\avg{C_{22}}$ becomes negligible at the same limit. Then, it is supposed that this behaviour is also valid under both the infinite Weissenberg and Reynolds number limits by taking carefully these limits, so that we go along the universal MDR asymptotic line. Hence, one might expect an intermediate region $\delta_\nu \ll y \ll \delta$ of approximately constant Reynolds shear stress, i.e. $-\avg{u'v'}/u_\tau^2 \to 1$, implied by the balance of shear stresses when taking the limits of $y/\delta \to 0$ and $y/\delta_\nu \to \infty$ with the reasonable assumption that $\nu\beta\frac{\dd}{\dd y}\avg{u} \to 0$ for $y \gg \delta_\nu$. In summary, the classical turbulence generation by $-\avg{u'v'}$ seems to survive at the MDR limit, based on the above assumptions.

Polymer-turbulence dynamical interactions were expressed through viscoelastic dissipation $\varepsilon_P \equiv \avg{\sigma'_{ij}\pd_{x_j}u'_i}$, which can either dissipate or produce turbulent kinetic energy. For HDR and MDR flows, $\int \varepsilon_P \,\dd y$ becomes vital in the flow dynamics in proportion to $\int \mathcal P \,\dd y$ and $\int \varepsilon_N \,\dd y$ due to the vast inhibition of Reynolds shear stress and fluctuating strain rates, respectively. In particular, a different view of the near-wall dissipation dynamics is shown for HDR/MDR flows, with the maximum dissipation arising away from the wall. It is intriguing to note that $\varepsilon_P$ follows a transitional pattern from LDR to HDR regime (see Fig. \ref{fig:energyterms}c) similar to $u'_+$ (see Fig. \ref{fig:velrms}a) and $\avg{C_{22}}$ (see Fig. \ref{fig:confcd}a). This characteristic behaviour is also reproduced on average in $\int \varepsilon_P \,\dd y$, where its dissipative feature enhances in the LDR regime but attenuates for HDR/MDR flows, with the productive nature dominating for high percentage drag reduction. Thus, polymers get somehow involved in the production dynamics of turbulent kinetic energy.

In view of the current viscoelastic DNS the following conceptual picture of drag reduction is deduced, which is an extension to and refinement of the mechanism proposed by Min \etal \cite{minetal03a}. Polymers in the near-wall region extract energy from the flow due to the uncoiling caused by the mean shear and release some portion of this stored elastic energy back to the flow by contracting as they move away from the wall. This transport of energy depends on Weissenberg number which determines the distribution of energy away from the wall. Ultimately, this process undermines the dynamics of near-wall turbulence. Note that polymers also unravel due to velocity fluctuations, as they move towards the core region of the flow, extracting again energy from the flow.
This mechanism appears to be valid for all drag reduction regimes with the dissipative and productive elements of viscoelastic dissipation competing in different parts of the flow for different levels of DR. We also observe that correlation $\avg{\sigma'_{1j}\pd_{x_j}u'_1}$ is able to resemble the dynamics of $\varepsilon_P$ and specifically that $\avg{\sigma'_{11}\pd_{x}u'}$ and $\avg{\sigma'_{13}\pd_{z}u'}$ are the correlations responsible for the production of turbulent kinetic energy by polymers. 

So far, in the limited context of the FENE-P model and at moderate Reynolds number DNS, the proposed phenomenology agrees with the majority of experimental and numerical data, where dampening of near-wall turbulence has long been speculated with various analyses and interpretations. Here, however, the transfer of energy from the flow to the polymers, its redistribution by the latter in the flow field and the prevalence of anisotropy over the components of $E \equiv \frac{1}{2}\avg{|\bm u'|^2}$ in the three Cartesian directions is suggested as a possible cause of drag reduction.

\begin{acknowledgments}
We are grateful to S. Laizet for providing the Navier-Stokes solver and to Halliburton for the financial support. We would also like to thank J. G. Brasseur, L. R. Collins and T. Vaithianathan for useful discussions.
\end{acknowledgments}
\appendix*
\section{Computational method}
\subsection{Numerical method for the FENE-P model}
\label{sec:fenepsolver}
\begin{figure}[!ht]
 \centering
 \includegraphics[width=8.5cm]{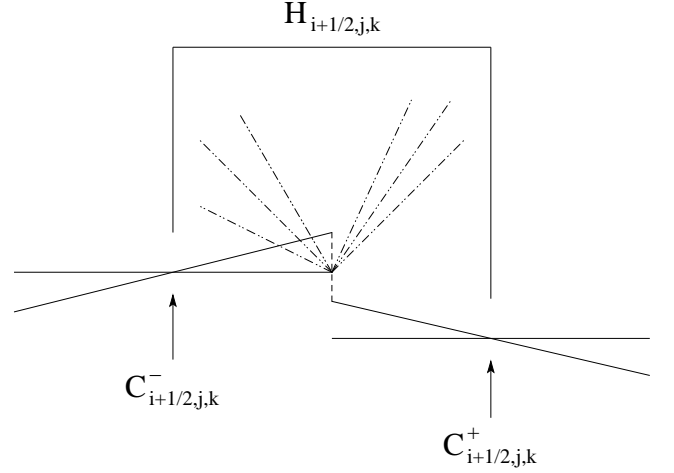}
 \caption{Central differencing approach -- staggered integration over a local Riemann fan denoted by the dashed-double dotted lines.}
 \label{fig:riemannfan}
\end{figure}

The numerical scheme adapted here for non-periodic boundary conditions was initially developed by Vaithianathan \etal \citet{vaithietal06} for periodic domains. The main idea behind the high-resolution central schemes employed here is the use of higher-order reconstructions, which enable the decrease of numerical dissipation so as to achieve higher resolution of shocks. In essence, they employ more precise information of the local propagation speeds. A key advantage of central schemes is that one avoids the intricate and time-consuming characteristic decompositions based on approximate Riemann solvers \citep{leveque02}. This is because these particular schemes realise the approximate solution in terms of its cell averages integrated over the Riemann fan (see Fig. \ref{fig:riemannfan}).

Considering the discretisation of the convection term of the FENE-P model only in the $x$-direction, using the reconstruction illustrated in Fig. \ref{fig:riemannfan}, the following second-order discretisation is obtained
\begin{equation}
 \frac{\pd \bm C_{i,j,k}^n}{\pd x} = \frac{1}{\D x}(H_{i+1/2,j,k}^n - H_{i-1/2,j,k}^n)
\end{equation}
where
\begin{align}
 \label{eq:Hplus}
 H_{i+1/2,j,k}^n & = \frac{1}{2} u_{i+1/2,j,k} (\bm C^+_{i+1/2,j,k} + \bm C^-_{i+1/2,j,k}) \nonumber\\
		 & - \frac{1}{2} |u_{i+1/2,j,k}| (\bm C^+_{i+1/2,j,k} - \bm C^-_{i+1/2,j,k})
\end{align}
with
\begin{equation}
 \bm C^\pm_{i+1/2,j,k} = \bm C_{i+1/2 \pm 1/2,j,k}^n
 \mp \frac{\D x}{2} \sdot \frac{\pd \bm C}{\pd x}\bigg|_{i+1/2 \pm 1/2,j,k}^n
\end{equation}
and
\begin{equation}
 \label{eq:slopelimiter}
 \frac{\pd \bm C}{\pd x}\bigg|_{i,j,k}^n = 
 \begin{cases}
  \frac{1}{\D x}(\bm C_{i+1,j,k}^n - \bm C_{i,j,k}^n) \\
  \frac{1}{\D x}(\bm C_{i,j,k}^n - \bm C_{i-1,j,k}^n) \\
  \frac{1}{2\D x}(\bm C_{i+1,j,k}^n - \bm C_{i-1,j,k}^n).
 \end{cases}
\end{equation}
Similarly, Eqs. \eqref{eq:Hplus}-\eqref{eq:slopelimiter} can be rewritten for $H_{i-1/2,j,k}^n$. The appropriate choice of the derivative discretisation in Eq. \eqref{eq:slopelimiter} limits the slope so that the SPD property for $\bm C$ is satisfied. The SPD criterion for this choice is that all the eigenvalues of the conformation tensor should be positive, viz. $\lambda_i > 0$ and subsequently all its invariants should be positive for at least one of the discretisations. Note that just $\det(\bm C) > 0$, is not sufficient to guarantee the SPD property for the tensor \citep{strang88}. In case none of the options in Eq. \eqref{eq:slopelimiter} satisfy the criterion, then the derivative is set to zero reducing the scheme to first order locally in space. The proof for $\bm C$ being SPD using this numerical scheme can be found in \citet{vaithietal06}. The eigenvalues of the conformation tensor in this implementation are computed using Cardano's analytical solution \citep{nr} for the cubic characteristic polynomial avoiding any complicated and time-consuming linear algebra matrix decompositions and inversions for just a $3 \times 3$ matrix. Ultimately, the advantage of this slope-limiter based method is that it adjusts in the vicinity of discontinuities so that the bounds on the eigenvalues cannot be violated, eliminating the instabilities that can arise in these types of calculations, without introducing a global stress diffusivity.

The complicated nature of the slope-limiting procedure raises difficulties in the case of wall boundaries for a channel flow computation, leading to loss of symmetry in the results. This had not been encountered by Vaithianathan \etal \citet{personalcomm07}, since they only considered periodic boundary conditions. So, the implementation of the numerical method near the walls of the channel was modified for this study considering ghost nodes beyond the wall boundaries to keep the original formulation unaltered, preserving in that way the second-order accuracy at the boundaries. The values at the ghost nodes were linearly extrapolated from the interior solution \citep{leveque02}, i.e.
\begin{align}
 \bm C_{i,j+1,k}^n &= \bm C_{i,j,k}^n + (\bm C_{i,j,k}^n - \bm C_{i,j-1,k}^n) \nonumber\\
 &= 2\bm C_{i,j,k}^n - \bm C_{i,j-1,k}^n.
\end{align}

The time advancement is done simply using the forward Euler update, treating implicitly the third, the forth term on the left hand side and the right hand side of Eq. \eqref{eq:nondimfenep} due to the potential finite extensibility of the polymer. Hence, the fully discretised form of the FENE-P model is
\begin{align}
 \label{eq:discretefenep}
 \bm C^{n+1}_{i,j,k} & = \bm C^{n}_{i,j,k} \nonumber\\
		     & - \frac{\D t}{\D x}(H^n_{i+1/2,j,k} - H^n_{i-1/2,j,k}) \nonumber\\
		     & - \frac{\D t}{\D y_j}(H^n_{i,j+1/2,k} - H^n_{i,j-1/2,k}) \nonumber\\
		     & - \frac{\D t}{\D z}(H^n_{i,j,k+1/2} - H^n_{i,j,k-1/2}) \nonumber\\
		     & + \D t (\bm C^{n+1}_{i,j,k} \grad \bm u^n_{i,j,k} 
		       + \grad \bm u^{n^\top}_{i,j,k} \bm C^{n+1}_{i,j,k}) \nonumber\\
		     & - \D t \lrbig{\frac{1}{\text{We}_c} f(\bm C^{n+1}_{i,j,k}) \bm C^{n+1}_{i,j,k} - \bm I}
\end{align}
with
\begin{align}
 \bm C^{n}_{i,j,k} = \frac{1}{6}( & \bm C^-_{i+1/2,j,k} + \bm C^+_{i-1/2,j,k} \nonumber\\
				+ & \bm C^-_{i,j+1/2,k} + \bm C^+_{i,j-1/2,k} \nonumber\\
				+ & \bm C^-_{i,j,k+1/2} + \bm C^+_{i,j,k-1/2} )
\end{align}
so that the convection term and the explicit term coming from the time derivative can be assembled in a convex sum
\begin{equation}
 \bm C^* = \bm C^n_{i,j,k} + \frac{\pd \bm C^n_{i,j,k}}{\pd \bm x} = \sum_{l = 1}^N s_l \bm C_l
\end{equation}
where all coefficients $s_l \geq 0$ satisfy $\sum_{l = 1}^N s_l = 1$, with $\bm C^*$ being SPD if the matrices $\bm C_l$ are SPD, ensuring the finite extensibility of the dumbbell, i.e. the trace of the conformation tensor is bounded $tr\bm C = \lambda_1 + \lambda_2 + \lambda_3 \leq L_P^2$ \citep{vaithietal06}. The following CFL condition needs to be satisfied for the coefficients $s_l$ to be non-negative
\begin{equation}
 \label{eq:cfl}
 \mathrm{CFL} = \max \lt\{ \frac{|u|}{\D x}, \frac{|v|}{\D y_{min}}, \frac{|w|}{\D z} \rt\} \sdot \D t < \frac{1}{6}
\end{equation}
and it also determines the time step $\D t$. Note that this CFL condition is more strict than the one for compact finite differences \citep{lele92} used for Newtonian turbulence computations.

The numerical solution of Eq. \eqref{eq:discretefenep} is carried out by first rewriting it in a Sylvester-Lyapunov form \citep{matrixcookbook}, separating the implicit and explicit terms, i.e.
\begin{equation}
 \label{eq:sylvester}
 \bm A^\top \bm X + \bm X\bm A = \bm B \Rightarrow (\bm I \otimes \bm A^\top + \bm A^\top \otimes \bm I) \bm x = \bm b
\end{equation}
where $\bm A \equiv \frac{1}{2}[1 + f(\bm C^{n+1}_{i,j,k})\frac{\D t}{\text{We}_c}]\bm I - \D t \grad \bm u^n_{i,j,k}$, $\bm X \equiv \bm C^{n+1}_{i,j,k}$ and $\bm B \equiv \bm C^* + \frac{\D t}{\text{We}_c}\bm I$ are $3 \times 3$ matrices, 
$(\bm I \otimes \bm A^\top + \bm A^\top \otimes \bm I)$ is a $9 \times 9$ matrix and $\bm x \equiv \mathrm{vec}(\bm X)$, $\bm b \equiv \mathrm{vec}(\bm B)$ are $9 \times 1$ vectors. The formula on the right hand side of Eq. \eqref{eq:sylvester} can be reduced from $9 \times 9$ to a $6 \times 6$ system of equations considering the symmetry of the conformation tensor. Note that Eq. \eqref{eq:sylvester} is non-linear and can now be solved using conventional methods. In this study, the Newton-Raphson method for non-linear systems was applied using the LU decomposition for the inversion of the Jacobian \citep{dennisschnabel83,nr}.
\subsection{Time advancement}
\label{sec:timeadvance}
After obtaining the new update of the conformation tensor $\bm C^{n+1}_{i,j,k}$, the two-step, i.e. three time-level, second-order Adams-Bashforth/Trapezoidal scheme is used for the time integration of Eqs. \eqref{eq:nondimNS2} through the following projection method \citep{peyret02}
\begin{equation}
 \frac{\bm u^\ast - \bm u^n}{\D t} = \frac{1}{2} (3\bm F^n - \bm F^{n-1}) + \frac{1}{2}(\bm P^\ast_{n+1} + \bm P^n)
\end{equation}
\begin{equation}
 \frac{\bm u^{n+1} - \bm u^\ast}{\D t} = - \grad\tilde p^{n+1}
\end{equation}
where
\begin{equation}
 \bm F = - \frac{1}{2} \lt[ \grad(\bm u \ox \bm u) + (\bm u \sdot \grad)\bm u \rt]
	+ \frac{1}{\text{Re}_c} \bm\D\bm u
\end{equation}
and
\begin{equation}
 \bm P = \frac{1-\beta}{\text{Re}_c\text{We}_c} \grad \sdot \lrbig{\frac{L^2_p - 3}{L^2_p - tr\bm C}\bm C - \bm I}
\end{equation}
with
\begin{equation}
 \tilde p^{n+1} = \frac{1}{\D t} \int_{t_n}^{t_{n+1}} p \, dt.
\end{equation}
The incompressibility condition $\grad \sdot \bm u^{n+1} = 0$ is verified by solving the Poisson equation
\begin{equation}
\label{eq:poisson}
 \grad \sdot \grad\tilde p^{n+1} = \frac{\grad \sdot \bm u^\ast}{\D t}
\end{equation}
which is done in Fourier space \cite{laizetlamballais09}. It is well known that these multistep methods are not self-starting and require a single-step method to provide the first time level \citep{peyret02,leveque07}. In this study, explicit Euler was chosen for just the first iteration of these computations, viz. $\bm u^n = \bm u^{n-1} + \D t \bm F^{n-1}$.
\bibliography{references}
\end{document}